\documentclass[11pt, prl,twocolumn, reprint,nofootinbib]{revtex4-1}
\usepackage{graphicx} 
\usepackage{amssymb}
\usepackage{amsmath}

\def\be{\begin{equation}}
\def\ee{\end{equation}}
\def\ba{\begin{eqnarray}}
\def\ea{\end{eqnarray}}
\def\p{\partial}
\def\la{~\mbox{\raisebox{-.6ex}{$\stackrel{<}{\sim}$}}~}
\def\ga{~\mbox{\raisebox{-.6ex}{$\stackrel{>}{\sim}$}}~}
\def\Z{{\mathbb Z}}

\begin{document}

\title{Natural Chaotic Inflation and UV Sensitivity}

\author{Nemanja Kaloper}
\affiliation{Department of Physics, University of California, Davis, Davis, CA 95616}
\author{Albion Lawrence}
\affiliation{Martin Fisher School of Physics, Brandeis University, Waltham, MA 02454}

\begin{abstract}
If the recent measurement of B-mode polarization by BICEP2 is due to primordial gravitational waves, it implies that inflation was driven by energy densities at the GUT scale $M_{GUT} \sim 2\times 10^{16} GeV$.  This favors single-field chaotic inflation models. These models require transplanckian excursions of the inflaton, forcing one to address the UV completion of the theory.  We use a benchmark 4d effective field theory of axion-4-form inflation to argue that inflation driven by a quadratic potential (with small corrections) is well motivated in the context of high-scale string theory models; that it presents an interesting incitement for string model building; and the dynamics of the UV completion can have observable consequences.

\vskip .2cm
\noindent{BRX-TH-676}
\end{abstract}

\maketitle

\section{Introduction}

The detection of B-mode polarization by the BICEP2 experiment \cite{Ade:2014xna,Ade:2014gua}\ gives tantalizing evidence that quantum gravity may be directly relevant for observational cosmology. If the future checks confirm that the BICEP2 B-modes are generated by primordial gravitational waves, this will give support to the idea that these waves are induced by quantum fluctuations of the graviton.
Furthermore, simple chaotic inflation models such as $V = \frac{1}{2} m^2 \phi^2$ \cite{Linde:1983gd} are in excellent current agreement with the data. To generate the observed density fluctuations in the CMB and large scale structure, chaotic inflation models occur at energy densities $\la M_{GUT}^4 \sim (2\times 10^{16}\ GeV)^4$. Any models generating observable primordial gravitational radiation require transplanckian excursions of the inflaton in field space of order $\Delta \phi \sim 10 m_{pl}$ \cite{Lyth:1996im, Efstathiou:2005tq}\ over the course of inflation.\footnote{We use the reduced Planck mass $m_{pl} = 2.4 \times 10^{18}\ GeV$; we take $M_{GUT}$ to be the value suggested by supersymetric coupling unification \cite{Dimopoulos:1981yj,Amaldi:1991cn,Langacker:1992rq}.} Therefore all such models are sensitive to the UV completion at the Planck scale, requiring a good understanding of quantum gravity.  The purpose of this paper is to consider constraints on the UV completion of models dominated by a (possibly distorted) quadratic potential, and present them as an incitement for string model builders.

Chaotic inflation is stable against perturbative quantum corrections, as exemplified as early as \cite{Smolin:1979ca,Linde:1987yb}, in response to the concerns about irrelevant operator contributions raised in \cite{Enqvist:1986vd}. This in fact follows if one protects the theory with (approximate) shift symmetries, so that all couplings to the inflaton are either very weak or via the derivatives of the inflaton. However, nonperturbative quantum-gravitational effects are expected to break such global symmetries, inducing ${\cal O}(1)$ coefficients for all Planck-suppressed operators.  When $\phi$ ranges over super-Planckian distances, these operators may spoil the small curvature of the inflaton potential required for slow roll inflation. Another concern is that in a string compactification, the degrees of freedom can shift substantially as fields move over super-Planckian distances; this can also manifest itself in dangerous Planck-suppressed operators. A good UV-complete realization must control these operators.\footnote{For a more complete review and discussion of the issues in this paragraph, see \cite{Kaloper:2011jz}.}

\begin{figure}
\includegraphics[scale=.75]{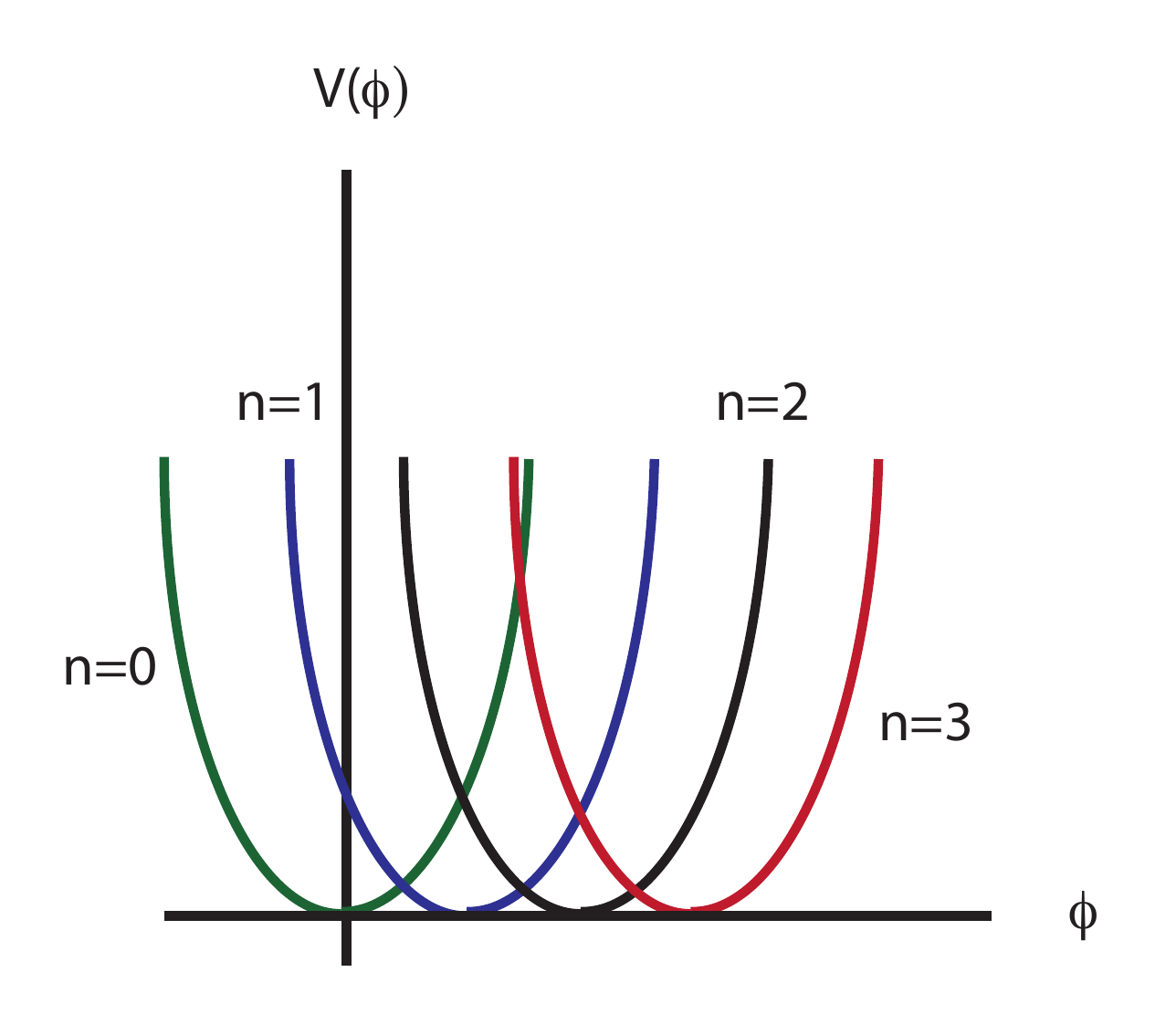}
\caption{Energies as a function of $\phi$, for the potential $V = \frac12 (\mu \phi + q)^2$. The picture repeats itself each time one shifts $\phi \to \phi + e^2/\mu \equiv \phi + f$.}
\end{figure}

Another technically natural high-scale model takes as the inflaton a periodic pseudoscalar (aka an axion) for which the potential is generated by instantons \cite{Freese:1990rb,ArkaniHamed:2003wu,ArkaniHamed:2003mz}: the topology of field space protects the shift symmetry of the inflaton.  The usual potential is $V \sim \Lambda^4 \cos(\phi/f)$.  However, high-scale slow roll inflation requires axion decay constants $f > \phi> m_{pl}$. Gravitational instantons, such as wormholes \cite{Kallosh:1995hi}, and string theoretic instantons \cite{Banks:2003sx}\ typically have actions of order $S \sim (\frac{m_{pl}}{f})^p$; thus higher-order instantons can spoil slow-roll inflation \cite{Banks:2003sx,ArkaniHamed:2006dz}.

Axion monodromy models control the Planck-suppressed operators by combining chaotic and natural inflation.  In this scenario, the inflaton is a compact scalar with periodicity $f < m_{pl}$. The presence of fluxes or branes ``unwraps" the inflaton configuration space \cite{Silverstein:2008sg,McAllister:2008hb,Kaloper:2008fb,Kaloper:2011jz,Dubovsky:2011tu}, leading to a multivalued potential as in Figure 1.  In what follows we will focus on the axion-4-form models of \cite{Kaloper:2008fb,Kaloper:2011jz}, which we dub ``natural chaotic inflation", as a benchmark theory from which to discuss the general phenomenon of axion monodromy. We start with the Lagrangian density
\be
	{\cal L} = \frac12 m_{pl}^2 R - \frac{1}{2} (\p\phi)^2 - \frac{1}{48} F_{(4)}^2 + \frac{\mu}{24} \phi {}^*F_{(4)}\, , \label{eq:axionfour}
\ee
where $F_{(4)} = dA_{(3)}$ is a four-form field strength, with $A_{(3)}$ a three-form gauge field.  The canonical momentum
$p_{A_{123}} = {}^*F_{(4)} - \mu\phi$ is quantized in units of the electric charge $e^2$; using this, one may write the Hamiltonian as
\be
	H = \frac{1}{2} \left(p_\phi\right)^2 + \frac{1}{2} \left(\nabla \phi\right)^2 + \frac{1}{2}\left(n e^2 + \mu \phi\right)^2 \, .\label{eq:afham}
\ee
If $\phi \to \phi + f$, we must have $\mu f = e^2$ for consistency.  The quantum number $n$ can be shifted by the nucleation of membranes, leading to the multivalued potential for $\phi$ shown in Figure 1.  At tree level, if membrane nucleation is suppressed, one has a model of chaotic inflation with a quadratic potential.  The fundamental periodicity of the scalar implies that corrections to $V = \frac{1}{2} \mu^2 \phi^2$ take the form $(V/M_{uv}^4)^n$, which can be small \cite{Kaloper:2008fb,Kaloper:2011jz}.  A danger to the model remains: couplings such as $\mu$ can depend on moduli with Planck-suppressed couplings. If the moduli masses are of the order or smaller than the Hubble scale $H$, they potentially destabilize inflation. 

The axion monodromy models of \cite{Silverstein:2008sg,McAllister:2008hb} turn these bugs into features, by providing a UV completion in which the overall effect of higher-order corrections and couplings to moduli is to flatten the tree-level potential \cite{Silverstein:2008sg,McAllister:2008hb,Dong:2010in}.  
The constructions in those papers give potentials significantly flatter than quadratic while still being large-field models, with significant tensor-to-scalar ratio $r$.  These may be consistent with the BICEP2 result, given the current statistical significance of the results. Further, additional UV complete models realizing the monodromy mechanism
starting from steeper potentials with larger $r$ are under development \cite{monopowers}. Here we will focus on whether a quadratic potential generated by integrating the 4-form starting with (\ref{eq:axionfour}) can be realized with only small corrections, using effective field theory as a guide.  
Our motivation for selecting this particular class of models is the relative simplicity of the dynamics which follows; yet we will find that there is a range of interesting potential signatures. 
We will discuss the relevant dynamical scales arising in various string theory scenarios, to argue that existing models consistent with grand unification are at the edge of being viable in this regard. If viable, they can give a finite probability for nonperturbative transitions to occur in early epochs of inflation \cite{Kaloper:2011jz}, and can give observable corrections to the tree-level quadratic potential.  In particular, we will show that the natural consequence of a single UV scale $M_{uv} \gtrsim M_{GUT}$ close to the GUT scale yields potentially observable corrections to the scalar and tensor spectrum. We present this as motivation for further work on high-scale string models.

\section{Scales and corrections}

The theory (\ref{eq:axionfour}) is protected from direct polynomial contributions to the axion potential of the for $\frac{\phi^{n}}{m_{pl}^{n-4}}$ by the periodicity of the scalar \cite{Kaloper:2008fb,Kaloper:2011jz}.  Corrections of the form 
\be
	\delta {\cal L} = \frac{c_n}{M_{uv}^{4n}} F^{2(n+1)} \, , \label{eq:fluxcorr}
\ee
for $n \ge 1$ are not forbidden.  Upon integrating out the four-form, these lead to corrections of the form
\be
	\delta V = c_n' V_{tree} \left(\frac{V_{tree}}{M_{uv}^4}\right)^n \, .
	\label{pertcorrs}
\ee
If $M_{uv} \gg M_{GUT}$ these are small.  They may become important if $M_{uv}$ is close to but still larger than $M_{GUT}$.  Note, that these corrections can be viewed as being generated by integrating out the gauge field $A$, whose field strength is $F = dA$.
In what follows we will focus mainly on the signatures of these corrections.

The other serious danger to this scenario (and to all models of high-scale inflation) arises from the coupling of moduli to the inflaton.  In models with geometric compactifications, moduli arise from the metric degrees of freedom which have kinetic terms weighted by $m_{pl}$.  Upon writing these degrees of freedom in terms of canonical scalars $\psi$, they appear in potential terms in the form $\psi/m_{pl}$.  We expect the moduli mass terms to be generated by potential energies of the form
\be
	W(\psi) = M^4 w\left(\frac{\psi}{m_{pl}}\right) \, , \label{eq:modpot}
\ee
where $M$ is the scale of the physics generating the moduli potential.  The masses are of the order $M_{mod}^2 = \frac{M^4}{m_{pl}^2}$.  During inflation, the inflaton potential provides an additional source term for the moduli,
\be
	{\cal V}(\phi,\psi) = \frac{1}{2} \mu^2\left(\frac{\psi}{m_{pl}}\right) \phi^2 + \frac{1}{2} M_{mod}^2 \psi^2 \, .
\ee
If $\p_{\psi}^2 V \sim \frac{V}{m_{pl}^2} \sim H^2 \gtrsim M_{mod}^2$, the modulus $\psi$ is potentially destabilized, absent some interesting mechanism; it can also lead to flattening of the potential as in \cite{Dong:2010in}. This can be avoided if $M \gg M_{GUT}$.\footnote{As discussed in \cite{Kaloper:2011jz}, this may not be sufficient if $\psi$ couples with more than gravitational strength, or if the minimum of $V$ with respect to $\psi$ is more than a Planck distance away from $\psi = 0$.} If $M \sim M_{uv}$, this will be achieved.

The theory may also have nonperturbative instabilities, generated by the discharge of the fluxes which give rise to the potential $V$.  There are two possibilities.  The first is that the inflaton jumps to a neighboring metastable branch of the multivalued inflaton potential, corresponding to a shift $n \to n -1$ in (\ref{eq:afham}).  These will be mediated by membranes charged under $A_{(3)}$.  As a guide, \cite{Kaloper:2011jz} found that if $f_{\phi} \sim 0.1 m_{pl}$, the transitions were suppressed if the charged membranes had tension $\sigma >  (0.2 V)^{3/4}$.  The second possibility is that a modulus $\psi$ controlling $\mu$ jumps, which is mediated by a nucleation of another membrane.  The space of possibilities is larger and depends on the jump in $\mu$ itself. In Ref. \cite{Kaloper:2011jz}, we estimated that for $\Delta\mu \sim 0.1\mu$, the transitions were suppressed if $\sigma > (0.8 V)^{3/4}$.  Note that if a transition of the former type does occur, one can realize the ``unwinding inflation" of \cite{DAmico:2012sz,DAmico:2012ji}: there are at least two directions in the configuration space of (\ref{eq:axionfour}) which potentially support inflation, and it would be worthwhile to study these directions in a single model.

In all these cases, if the ``thin wall" approximation holds, we found that the transition probability is well approximated by the flat space result \cite{Kobzarev:1974cp,Coleman:1977py}:
\be
	P \sim \exp\left(- \frac{27 \pi^2 \sigma^4}{2 (\Delta V)^3}\right) \, . \label{bubprob}
\ee
For jumps in $n$, $\Delta V \sim \frac{\phi}{f} V$, while for jumps in $\mu$, $\Delta V \sim \frac{\Delta \mu}{\mu} V$.  This means that such processes are increasingly disfavored at later stages of inflation. However the probability is larger at larger values of $\phi$, implying that the final stage of smooth slow roll with small perturbations is of limited duration, as in unwinding inflation \cite{DAmico:2012sz,DAmico:2012ji}. Such transitions could generate interesting signatures early on, or equivalently on the largest visible scales on the sky.

\section{A sampling of string models}

A high scale of inflation $V\sim M_{GUT}^4$ presents a healthy incitement to string model builders.  Some attractive models of axion monodromy inflation, both global compactifications and ``local" models describing a patch of a putative compactification, have been constructed in \cite{Silverstein:2008sg,McAllister:2008hb,Dong:2010in,Dubovsky:2011tu}.  These describe models with a potential flattened away from the quadratic minimum.

To date there has been no deep investigation of models realizing the effective field theory described in \cite{Kaloper:2008fb,Kaloper:2011jz}, though a schematic version was described in \cite{Kaloper:2008fb}.\footnote{Papers \cite{Shlaer:2012by,Palti:2014kza} also present a string construction realizing a quadratic potential, although there is not a complete compactification scenario there.} We are interested in:

\begin{itemize}
\item the scale $M_{uv}$ which controls the corrections in Eq. (\ref{eq:fluxcorr}), usually corresponding to the fundamental string or 10/11d Planck scale. Our scenario requires $M_{uv} \gg M_{GUT}$.
\item The scale $M$ which controls the potential for moduli with generic couplings to the inflation; to avoid excessive corrections to our tree-level action, we are interested in whether $M \gg M_{GUT}$ for these moduli.
\end{itemize}
The additional challenge is that these models must also have a somewhat lower scale $\mu \sim 0.1 M_{GUT}^2/m_{pl}$ controlling the tree-level inflaton potential.

In order to gain some intuition, let us consider two scenarios broadly consistent with grand unification at the scale $M_{GUT}$, with candidates for realizing our axion-four form model. A review of the dynamical scales in string models consistent with grand unification (with the aim of looking for other signatures of high-scale physics in the CMB) can be found in \cite{Kaloper:2002uj}.

One example is M theory compactified on a manifold of $G_2$ holonomy, following \cite{Harvey:1995ne,Acharya:1996ci,Witten:2001uq,Acharya:2001gy,Witten:2001bf},.These models are described by a $K3$ fibration over a base $S^3/\Z_n$, whose size sets the GUT scale.  
The existence of axion-four form monodromy was pointed out in \cite{Beasley:2002db}; for an alternate discussion in the case of toroidal compactifications of M-theory, which captures the essential physics, see \cite{Kaloper:2008qs,Kaloper:2008fb}.  
If the GUT theory is realized following \cite{Harvey:1995ne,Acharya:1996ci,Witten:2001uq,Acharya:2001gy,Witten:2001bf}, consistency with grand unification leads to \cite{Kaloper:2002uj}:
\be
	m_{11} \sim M_{GUT}\, , ~~~~~  \ \ L_{K3}^{-1} \sim 0.1 M_{GUT} \, .
\ee
We expect a typical modulus potential term of the form (\ref{eq:modpot}) will have $M$ at most equal to $m_{11}$, up to pure numerical factors, while $M_{uv}$ in (\ref{eq:fluxcorr}) should also be of this order. In this case, our scenario looks tenuous.

We can also consider type IIB theories.  We leave explicit realizations of axion-four form monodromy for future work, and simply discuss the dynamical scales of this class of models. 
In typical models consistent with grand unification at $M_{GUT}$, the realization of the GUT scale will depend on the origins of the gauge degrees of freedom.  If the gauge group lives on D3-branes, the gauge coupling will be the string coupling -- thus the latter is weak.  The natural scale for the GUT symmetry to be broken is the string scale, in which case $M_{uv}^4 \sim V$ and corrections of the form (\ref{eq:fluxcorr}) are problematic.  Furthermore, one is left with a Kaluza-Klein scale that is of order $0.1 M_{GUT}$.  Following the discussion in \cite{Kachru:2006em}, we find that the heaviest moduli, stablized by three-form flux, have masses of order $H$.  If one instead realizes the gauge group on D7-branes wrapping finite-sized four-cycles, and sets $M_{GUT}$ to be equal to the Kaluza-Klein scale and the string coupling to be ${\cal O}(1)$ (appropriate since D7-brane source a dilaton), the string scale is of order $2 M_{GUT}$. If $M_{uv} = m_s$, these models are just at the edge of viability from the standpoint of our scenario of inflation.  Again following the discussion in \cite{Kachru:2006em}, moduli stabilized by closed string or brane worldvolume fluxes will have masses of order $M_{uv}^2/m_{pl}$.  However, other moduli are often parametrically lighter, so the inflaton must be shielded from these. We note that this discussion is consistent with the explicit constructions given in \cite{Cicoli:2012vw,Cicoli:2013mpa,Cicoli:2013cha}, in cases when the string coupling is pushed towards the GUT scale.

We stress that the examples given here are meant to be an illustration.
Our point is simply that for string theory to produce a model with a quadratic potential for the inflaton, which is  consistent with the existing CMB data, it is natural to consider compactifications a fairly high scale, close to the higher-dimensional Planck scale; furthermore one must work hard to ensure that the moduli do not spoil the scenario.  
It is not clear that geometric compactifications are the right arena to work in. Note also that a rescaling of $M_{uv}$, $M$ by a factor of a few can make a significant difference. As we will see below, a consequence is that such models will fairly easily produce observable corrections of the form (\ref{pertcorrs}). It is not clear how much precise quantitative control we can gain over them in the near future.  However, we advocate that the problem of making some moduli heavy and sequestering the inflaton from the remaining moduli, in a class of high scale models, deserves further scrutiny in and of itself. 

One should also pay additional attention to how the inflaton couples to other fields to ensure that reheating can in principle proceed smoothly without overproducing long-lived moduli. Some phenomenological explorations from the bottom up, such as \cite{Hall:2014vga}, may also provide interesting points of contact with the high energy theory.

\section{Observable consequences of corrections}

The scalar and tensor spectra for single field slow roll inflation are, respectively.\footnote{Here we use the conventions of \cite{Baumann:2008aq}, which differ from those we used in \cite{Kaloper:2011jz}.}
\be
	P_{\tt S} = \Bigl( \frac{H^2}{2\pi \dot \phi} \Bigr)^2 = \frac{1}{24\pi^2 m_{pl}^4} \frac{V}{ \epsilon}\
	\label{spectras}
\ee
and
\be	
	P_{\tt T} = \frac{8}{m_{pl}^2} \Bigl(\frac{H}{2\pi}\Bigr)^2 =\frac{2}{3\pi^2} \frac{V}{m_{pl}^4}  \, .
	\label{spectra}
\ee
The tensor-to-scalar ratio is
\be
	r = 16 \epsilon = \frac{8}{m_{pl}^2} \Bigl(\frac{\dot \phi}{H}\Bigr)^2 \, ,	\label{scaltenr}
\ee
where $\epsilon =  \frac{1}{2m_{pl}^2} \Bigl(\frac{\dot \phi}{H}\Bigr)^2$ is a slow roll parameter. When discussing explicit numbers, we use the following observed values as constraints:
\begin{eqnarray}
	\sqrt{P_{\tt S}} & \sim & 5 \times 10^{-5} \, , \nonumber\\
	r & \sim & 0.2^{+.07}_{-.05}\ {\rm without\ dust\ subtraction} \, , \nonumber\\
	& \sim & 0.16^{+.06}_{-.05}\ {\rm with\ dust\ subtraction} \, , \label{eq:ourdata}
\end{eqnarray}
computed at 50 efoldings before the end of inflation. The power spectrum is based on the COBE result; the tensor-to-scalar ratios come from the BICEP2 result.
We will compute corrections to the tensor and scalar spectrum as a parametric function of the number of efolds left before the end of inflation ${\cal N}$ (which serves direclty as the clock) and the corrections to $V$; to estimate the size we will use the numbers above. Future data could change the observed value of $r$, and our choice of ${\cal N} = 50$ implies a choice of reheating history. This will change the detailed numbers we calculate below, but not our qualitative conclusions.\footnote{Note that corrections to ${\cal N}$-independent quantities used in \cite{Creminelli:2014oaa}\ are parametrically of the same order as the corrections to the spectrum we find here.}

First, we review the simple tree-level quadratic potential. The number of efolds before the end of inflation is ${\cal N} = \frac{1}{4} \Bigl(\frac{\phi}{m_{pl}}\Bigr)^2$.  The resulting scalar and tensor power are
\be
	P_{\tt S} = \frac{\mu^2}{6\pi^2 m_{pl}^2}{\cal N}^2 \, , ~~~~~~~~
	P_{\tt T} =\frac{4\mu^2}{3\pi^2 m_{pl}^2} {\cal N}  \, ,
	\label{spectraclock}
\ee
so that $r = \frac{8}{\cal N}$. The observed value of $P_S$ and ${\cal N} = 50$ fixes the inflaton mass 
\be
	\mu \simeq 1.8 \times 10^{13} GeV \, .
	\label{mass}
\ee
From this it follows that the tensor-to-scalar ratio is $r = 0.16$, and the energy scale of inflation is
$V^{1/4}_{50} = (2 \mu^2 m_{pl}^2 {\cal N}_{50})^{1/4} = 2.1 \times 10^{16} GeV$. At 60 efolds before the end, the scale of inflation is 
$V^{1/4}_{60} = 2.2 \times 10^{16} GeV$, not much different. Thus the scale of quadratic inflation is {\it very close} to the GUT scale $M_{GUT} \sim 2  \times 10^{16} GeV$. This is consistent with the BICEP2 result.

The principal sources of corrections to this scenario arise from membrane nucleation, corrections of the form (\ref{eq:fluxcorr}), and light moduli \cite{Kaloper:2008fb,Kaloper:2011jz}.\footnote{We will put aside for now the interesting possible corrections arising from periodic modulations of the potential \cite{Silverstein:2008sg,McAllister:2008hb,Kaloper:2011jz}.} First, consider nucleation of a bubble which changes $\mu$ by $\Delta\mu \sim 0.1 \mu$. Suppression of this effect (to preserve large scale homogeneity and isotropy) leads to a bound $\sigma^{1/3} \ga 9.8 \times 10^{15} GeV$ on the membrane tension, very close to the GUT scale.  Since $\Delta V = 2 (\Delta \mu/\mu) V$ in this case, the probability will increase in earlier epochs.  How early that occurs is exquisitely sensitive to the scale setting the membrane tension, due to the large powers which appear in (\ref{bubprob}). For example, if the membrane tension is precisely $M_{GUT}^3$, and $\Delta \mu \sim 0.1 \mu$, the nucleation probability is of order $1/e$ when $N \sim 960$, $V \sim 3.6 \times 10^{16}\ GeV$, $\phi \sim 44 m_{pl}$. A slight decrease in $\sigma^{1/3}$ places this epoch close to the observable one.
Thus, if the relevant scales are close to the GUT scale, our universe may have been the result of a bubble nucleation in a relatively near past. This can lead to a small negative spatial curvature \cite{Freivogel:2005vv,mattlenny}. Moreover, the bubble walls themselves can fluctuate, leading to a deformation of the homogeneous slices inside the bubble.  If one assumes that the tension is set by the GUT scale, that small fluctuations are governed by the action $S_{bubble} = \sigma \int d^3 x (\p X)^2$, and that $\delta (\sqrt{\sigma} X) \sim \sqrt{H}$, the fluctuations of these homogenous slices are comparable to the fluctuations due to the inflaton.  The authors of \cite{DAmico:2013iaa} have argued that this could yield a large scale asymmetry in the power spectrum, in line with claims that such an asymmetry has been observed in the CMB \cite{Eriksen:2003db}.

The corrections (\ref{pertcorrs}) to the tree level potential (\ref{eq:afham}) can affect the slow roll regime during the observable epoch of inflation in a significant way if $M_{uv} \gtrsim M_{GUT}$.
As we have discussed, this occurs for many string models consistent with grand unification. 
Because of the particular nature of the our baseline tree level quadratic potential, the slow roll corrections are particularly simple, being parameterized by $\epsilon \sim 0.01$.  We will find that that the new physics corrections (\ref{pertcorrs}) give significantly larger corrections.
Since $V_{tree} \sim M_{GUT}^4$, the corrections can be comparable to the tree level potential. For example, the first term in the expansion induces a quartic coupling, $c_1' V_{tree} \frac{V_{tree}}{M_{uv}^4} = \lambda_4 \phi^4/4$ with the induced coupling
\be
	\lambda_4 = 4 c_1' (\frac{\mu}{M_{uv}})^4 \, . 
\ee
Using the mass of the inflaton (\ref{mass}) and taking $M_{uv} \ga M_{GUT} =  2 \times 10^{16} GeV$, this yields 
\be
	\lambda_4 \la 6.6\times 10^{-13} \,  c_1'   \, .
\ee
where we expect $c_1' \sim {\cal O}(1)$.  A purely quartic inflaton potential consistent with the observed $P_{\tt S}$ has $\lambda \sim 2.8 \times 10^{-13}$.  Such a potential does not fit current values of $n_s$ and $r$: it must be subdominant to the quadratic term.  However, since $\lambda_4$ scales as $1/M_{uv}^{4}$, a slight increase in $M_{uv}$ can place $\lambda$ in a range consistent with current observations.  As $M_{uv}$ can be within a factor of a few of $M_{GUT}$, our theory is at the edge of respectability \cite{Kaloper:2011jz}. It is therefore important to examine these corrections carefully: they can correct the amplitude of the scalar density spectrum generated by the tree level quadratic potential by tens of percents \cite{Linde:1983gd}. 

We assume the corrections are subleading, so that the Taylor expansion for the effective potential whose terms are given in (\ref{pertcorrs}) is well within its radius of convergence: upon fitting to (\ref{eq:ourdata}), this assumption will turn out to be self-consistent. Consider a quadratic potential with a dominant subleading term $\Delta V = c_n' V_{tree} \Bigl(\frac{V_{tree}}{M_{uv}^4}\Bigr)^n = \frac{\lambda_{2(n+1)}}{2(n+1)} \phi^{2(n+1)}$, where 
$\lambda_{2(n+1)} = 2(n+1) c_n' \frac{\mu^{2(n+1)}}{M_{uv}{}^{4n}} = 2(n+1) \hat c_n\frac{\mu^{2(n+1)}}{M_{GUT}{}^{4n}} $, and $\hat c_n = c_n' (\frac{M_{GUT}}{M_{uv}})^{4n}$. 
We compute the corrections to the inflationary dynamics (in the slow roll approximation) to the first nontrivial order in $\lambda_{2(n+1)}$. We expect that the lowest values of $n$ will dominate the corrections to inflationary observables. First, if $V_{tree}/M_{uv}^4 < 1$, higher powers are small to begin with.  Secondly, corrections from higher powers die off more quickly during inflation: between ${\cal N}_2$ and ${\cal N}_1$ efoldings of inflation, the $n$th term will reduce by a factor of $\sim {\cal O}\left(({\cal N}_1/{\cal N}_2)^{n+1}\right)$.

We wish to express physical quantities during inflation in terms of the number ${\cal N}$ of efoldings before the end of inflation.  Since 
\be
	{\cal N} = \int^{\phi}_{\phi_{end}} \frac{d\phi}{m_{pl}^2} \frac{V}{\partial_\phi V} \, ,
	\label{newclock}
\ee
using $V = V_{tree} + \Delta V$ and expanding in $\Delta V$ we first find a correction\footnote{In what follows we will concentrate on monomial corrections for simplicity. Since we work to linear order in $\lambda_{2(n+1)}$, more complicated cases with corrections involving several different low powers can be obtained by superposition, taking care with possible cross-contamination of different terms by slow roll corrections.}
\be	
	{\cal N} = \frac14 \Bigl(\frac{\phi}{m_{pl}}\Bigr)^2 - \frac{n \lambda_{2(n+1)}}{4(n+1)^2} \frac{\phi^{2(n+1)}}{m_{pl}^2 \mu^2} \, , 
	\label{eq:Nfromphi}
\ee
where we ignore the ${\cal O}(1) \ll {\cal N}$ contribution of $\phi_{end}$ in (\ref{newclock}). Inverting (\ref{eq:Nfromphi}), we find:
\be
	 \frac14 \Bigl(\frac{\phi}{m_{pl}}\Bigr)^2 = {\cal N} + 2^{2n+1}  \frac{n \hat c_n}{(n+1)} \Bigl(\frac{\mu m_{pl}}{M_{GUT}{}^2}\Bigr)^{2n} {\cal N}^{n+1} \, .
	 \label{phiN}
\ee
At the order we are working in, we can insert the tree-level expression for $\mu$ into the correction term of (\ref{phiN}), to find
$ \frac14 \Bigl(\frac{\phi}{m_{pl}}\Bigr)^2 = {\cal N} \Bigl[1 + 2 \frac{n \hat c_n}{(n+1)} \Bigl(\frac{\cal N}{21}\Bigr)^n \Bigr]$. For the correction to be subdominant for ${\cal N} < 60$, $\hat c_n < \frac{n+1}{2n} (0.35)^n$. 

Next, the corrected values of the potential $V$ and the slow roll parameter $\epsilon$ in terms of ${\cal N}$ are\footnote{The factor of $21$ and the prefactor in $V$ arises from (\ref{mass}), which is slightly dependent on ${\cal N}$. One may replace ${\cal N}/21 \to (2 \mu m_{pl}/M_{GUT}^2)^2 {\cal N}$ in the formulae below for a more general expression, but the difference will be down by a factor of $\epsilon$.}
\ba
	V &=&  2 \mu^2 m_{pl}^2 {\cal N} \Bigl( 1 + \frac{2^{2n+1}(2n+1)}{n+1} \hat c_n \left(\frac{\mu m_{pl}}{M_{GUT}{}^{2}}\right)^{2n} {\cal N}^n\Bigr) \, \nonumber \\
	&=& \Bigl(0.45 \cdot 10^{16} GeV\Big)^4 {\cal N} \Bigl(1 + \frac{2(2n+1)}{n+1} \hat c_n (\frac{\cal N}{21})^n \Bigr)\, , \nonumber \\
	\epsilon 
        &=&  \frac{1}{2{\cal N}} \Bigl(1 +  \frac{2n(2n+1)}{n+1} \hat c_n (\frac{\cal N}{21})^{n}\Bigr)	\, .
	 \label{Vepsiloncorr}
\ea
Using Eqs. (\ref{spectras}), (\ref{spectra}) we obtain the corrected formulas for the scalar and tensor spectra, 
\ba
	P_{\tt S} 
	&=& \frac{\mu^2}{6\pi^2 m_{pl}^2}{\cal N}^2  \Bigl( 1 - \frac{2(2n+1)(n-1)}{n+1} \hat c_n (\frac{\cal N}{21})^n\Bigr)  \, , \nonumber \\
	P_{\tt T} 
	&=& \frac{4\mu^2}{3\pi^2 m_{pl}^2} {\cal N} \Bigl( 1 + \frac{2(2n+1)}{n+1}\hat c_n (\frac{\cal N}{21})^n\Bigr)  \, , 
	\label{spectraclockcorr}
\ea
A clear effect of the corrections is to enhance (for ${\hat c}_n > 0$) 
the tensor-to-scalar ratio relative to that generated by the tree-level potential, $r_{tree~50} = 0.16$,
\be
	r = 16 \epsilon =  \frac{8}{{\cal N}} \Bigl(1 +  \frac{2n(2n+1)}{n+1} \hat c_n (\frac{\cal N}{21})^{n}\Bigr) \, .
	\label{ratiocorr}
\ee

We claimed above that we can ignore corrections at high orders in $n$.  If we demand that the corrections be small, we must have $\hat c_n < \frac{n+1}{2n(2n+1)}(0.32)^n$. Since ${\hat c}_n \sim (M_{GUT}/M_{uv})^{4n}$, this power law behavior is entirely plausible if $M_{uv} > 1.3 M_{GUT}$. Corrections then appear as a series in ${\hat c}_n ({\cal N}/21)^n \sim (0.76)^n$ even in this worst case scenario.

The corrections at $n \sim {\cal O}(1)$ can be significant in early epochs of inflation, and larger than the corrections to slow roll. For a quadratic potential, the latter are multiplicative corrections of order $\epsilon, \eta \sim 0.01$.
Let us consider the cases of $n=1, 2$ corresponding to quartic and sextic corrections to the inflaton potential. In the quartic case, if we fit the potential to $r = 0.2$ at ${\cal N} = 50$, we find that ${\hat c}_1 \sim 0.035$, corresponding to $M_{uv} \sim 2.3 M_{GUT}$. At the earlier epoch ${\cal N} = 60$, such a correction gives multiplicative corrections of order $0.3$ the tree level expressions for $\epsilon$  $V$. Note that the correction to $P_{\tt S}$ vanishes at this order for a quartic correction. For a sextic correction, setting , $r = 0.2$ at ${\cal N} = 50$ implies ${\hat c}_2 \sim 0.0066$.  At ${\cal N} = 60$, this gives multiplicative corrections of order $0.36$ to $\epsilon$, and of order $0.18$ to $V$.  In particular, note that $\Delta P_{\tt S}/P_{\tt S} \sim - 0.18$ (if $c_1'$ is positive) at this order. 

The possible reduction of ${P}_{\tt S}$ induced by $n \geq 2$ terms is interesting given the claimed observation of the lack of power at low 
multipoles \cite{Eriksen:2003db}. Obviously, higher order corrections can help account for some of it, suppressing the scalar spectrum power by slightly increasing the inflaton rolling. 
Other mechanisms involving additional light fields that induce a small amount of curvaton-induced perturbations 
have also been proposed, see \cite{sloth}.  In fact such a scenario might occur in the presence of any additional light moduli in our framework, which couple to the inflaton only weakly, or not at all. 
 
The spectral indices $n_{\tt S} = 1 + \frac{d \ln P_{\tt S}}{d \ln k} =  1 - \frac{1}{P_{\tt S}} \frac{dP_{\tt S}}{d{\cal N}}$ and $n_{\tt T} = \frac{d \ln P_{\tt T}}{d \ln k} = - \frac{1}{P_{\tt T}} \frac{dP_{\tt T}}{d{\cal N}}$ are a useful diagnostic of corrections to $V_{tree}$.
From 
(\ref{spectraclockcorr}) we find:
\ba
	n_{\tt S} 
	&=& 1 - \frac{2}{\cal N} +  \frac{2n(2n+1)(n-1)}{21(n+1)} \hat c_n \Bigl(\frac{\cal N}{21}\Bigr)^{n-1} \, , \nonumber \\
	n_{\tt T} 
	&=& - \frac{1}{\cal N}  - \frac{2 n(2n+1)}{21(n+1)} \hat c_n \Bigl(\frac{\cal N}{21}\Bigr)^{n-1} = - 2\epsilon \, .
	\label{spectraindex}
\ea
The last of these two equations is the inflationary consistency condition \cite{Liddle:1992wi}, $n_{\tt T} = - 2\epsilon = - r/8$, which must be true for single field slow roll inflation. Further, the subleading corrections to $n_{\tt S}$ and $n_{\tt T}$, are related, their ratio being fixed by the power of the dominant correction $2(n+1)$. In case when $n=1$ and so $\Delta V$ is quartic, the corrections to the spectral indices are scale invariant -- that is, independent of ${\cal N} \propto k$. This should be expected since the quartic potential is scale invariant. In this case, the quantitative corrections obey $\Delta n_{\tt S} = 0$, $\Delta n_{\tt T} =  - 3 \hat c_1/21 \la 10^{-2}$, if we fix the quartic correction by demanding $r = 0.2$. A combination of $r, n_S$ can be formed which isolates corrections to the quadratic potential \cite{Creminelli:2014oaa}.  If possible, it will also be important to measure $n_T$ to test the consistency condition, as recently stressed in \cite{Dodelson:2014exa}.
	 
More generally, one should also consider the case where the subleading corrections to the potential 
compete directly with the quadratic potential in the earliest epochs of inflation, to see their effect on the data. This is a natural possibility if, as we stated, the UV completion is near the GUT scale. The quadratic potential eventually dominates, but higher-order corrections can exert more influence on the largest scales on the sky. A detailed fitting of the data is beyond the scope of our work here; some attempts fitting polynomial potentials may be found in \cite{Kobayashi:2014jga}.

The bottom line is that our examples demonstrate that subleading corrections are observationally relevant (and not yet ruled out) when the leading potential is quadratic and the UV cutoff of the low energy effective theory governing inflation is close to $M_{GUT}$.

\section{Summary -- Chaotic Evil Inflation}

The observation of primordial gravitational waves by BICEP2 has ushered a new era in cosmology. The results are pointing very strongly to the large field slow roll models of inflation  \cite{Linde:1983gd} as the dynamics that shaped the universe (although a lively discourse on this is still ongoing, 
\cite{Collins:2014yua,Contaldi:2014zua,Kehagias:2014wza,Dent:2014rga,Freese:2014nla,Czerny:2014qqa,Lyth:2014yya,DiBari:2014oja,Creminelli:2014oaa,Ashoorioon:2014nta,Ho:2014xza}). This presents a constructive incitement to model building, in particular in string theory, because large field models are sensitive to the details of the UV completion in which they should be embedded. 

In this work we have addressed the UV sensitivity of the specific class of models described in \cite{Kaloper:2008fb,Kaloper:2011jz}, which lead to quadratic potentials and are in excellent agreement with the current cosmological data\footnote{After the appearance of the first version of this work, several papers have appeared with very similar constructions of effective large field models of inflation from string theory \cite{Marchesano:2014mla,Hebecker:2014eua,Blumenhagen:2014gta,Harigaya:2014eta,Grimm:2014vva,Ibanez:2014kia}.}. Since such models operate at the GUT scale, and string theory consistent with grand unification have fundamental scales close to the GUT scale, the UV physics may leave imprints in the sky in the form of small corrections to the leading order CMB observables. These signatures can at least be used to constrain the UV physics. Even more interestingly, future observational tests could use these signatures as benchmarks to search for, in order to further test large field inflation.

We feel the discussion here only reinforces the conclusion of \cite{Kaloper:2011jz}: that this class of chaotic inflation models lives at the edge of respectability, which is often the most interesting place to be.

\vskip.5cm

{\bf Acknowledgments}: 
We would like to thank Guido D'Amico, Matt Kleban, Andrei Linde, Fernando Quevedo, Eva Silverstein, Martin Sloth, Lorenzo Sorbo and Alexander Westphal for very useful discussions. AL would like to thank the Aspen Center for Physics for its hospitality during the ``New Perspectives on Thermalization" winter conference, at the onset of this work. NK is supported by the DOE Grant DE-FG03-91ER40674. AL is supported by the DOE grant DE-SC0009987. 

\bibliography{axmod_refs.bib}

\begin{thebibliography}{69}%
\makeatletter
\providecommand \@ifxundefined [1]{%
 \@ifx{#1\undefined}
}%
\providecommand \@ifnum [1]{%
 \ifnum #1\expandafter \@firstoftwo
 \else \expandafter \@secondoftwo
 \fi
}%
\providecommand \@ifx [1]{%
 \ifx #1\expandafter \@firstoftwo
 \else \expandafter \@secondoftwo
 \fi
}%
\providecommand \natexlab [1]{#1}%
\providecommand \enquote  [1]{``#1''}%
\providecommand \bibnamefont  [1]{#1}%
\providecommand \bibfnamefont [1]{#1}%
\providecommand \citenamefont [1]{#1}%
\providecommand \href@noop [0]{\@secondoftwo}%
\providecommand \href [0]{\begingroup \@sanitize@url \@href}%
\providecommand \@href[1]{\@@startlink{#1}\@@href}%
\providecommand \@@href[1]{\endgroup#1\@@endlink}%
\providecommand \@sanitize@url [0]{\catcode `\\12\catcode `\$12\catcode
  `\&12\catcode `\#12\catcode `\^12\catcode `\_12\catcode `\%12\relax}%
\providecommand \@@startlink[1]{}%
\providecommand \@@endlink[0]{}%
\providecommand \url  [0]{\begingroup\@sanitize@url \@url }%
\providecommand \@url [1]{\endgroup\@href {#1}{\urlprefix }}%
\providecommand \urlprefix  [0]{URL }%
\providecommand \Eprint [0]{\href }%
\@ifxundefined \urlstyle {%
  \providecommand \doi  [0]{\begingroup \@sanitize@url \@doi}%
  \providecommand \@doi [1]{\endgroup \@@startlink {\doibase
  #1}doi:\discretionary {}{}{}#1\@@endlink }%
}{%
  \providecommand \doi  [0]{doi:\discretionary{}{}{}\begingroup
  \urlstyle{rm}\Url }%
}%
\providecommand \doibase [0]{http://dx.doi.org/}%
\providecommand \Doi [0]{\begingroup \@sanitize@url \@Doi }%
\providecommand \@Doi  [1]{\endgroup\@@startlink{\doibase#1}\@@Doi}%
\providecommand \@@Doi [1]{#1\@@endlink}%
\providecommand \selectlanguage [0]{\@gobble}%
\providecommand \bibinfo  [0]{\@secondoftwo}%
\providecommand \bibfield  [0]{\@secondoftwo}%
\providecommand \translation [1]{[#1]}%
\providecommand \BibitemOpen [0]{}%
\providecommand \bibitemStop [0]{}%
\providecommand \bibitemNoStop [0]{.\EOS\space}%
\providecommand \EOS [0]{\spacefactor3000\relax}%
\providecommand \BibitemShut  [1]{\csname bibitem#1\endcsname}%
\bibitem [{\citenamefont {Ade}\ \emph {et~al.}(){\natexlab{a}}\citenamefont
  {Ade} \emph {et~al.}}]{Ade:2014xna}%
  \BibitemOpen
  \bibfield  {author} {\bibinfo {author} {\bibfnamefont {P.}~\bibnamefont
  {Ade}} \emph {et~al.} (\bibinfo {collaboration} {BICEP2 Collaboration}),\
  }\href@noop {} { ({\natexlab{a}})},\ \Eprint {http://arxiv.org/abs/1403.3985}
  {arXiv:1403.3985 [astro-ph.CO]} \BibitemShut {NoStop}%
\bibitem [{\citenamefont {Ade}\ \emph {et~al.}(){\natexlab{b}}\citenamefont
  {Ade} \emph {et~al.}}]{Ade:2014gua}%
  \BibitemOpen
  \bibfield  {author} {\bibinfo {author} {\bibfnamefont {P.~A.~R.}\
  \bibnamefont {Ade}} \emph {et~al.} (\bibinfo {collaboration} {BICEP2
  Collaboration}),\ }\href@noop {} { ({\natexlab{b}})},\ \Eprint
  {http://arxiv.org/abs/1403.4302} {arXiv:1403.4302 [astro-ph.CO]} \BibitemShut
  {NoStop}%
\bibitem [{\citenamefont {Linde}(1983)}]{Linde:1983gd}%
  \BibitemOpen
  \bibfield  {author} {\bibinfo {author} {\bibfnamefont {A.~D.}\ \bibnamefont
  {Linde}},\ }\Doi {10.1016/0370-2693(83)90837-7} {\bibfield  {journal}
  {\bibinfo  {journal} {Phys. Lett.},\ }\textbf {\bibinfo {volume} {B129}},\
  \bibinfo {pages} {177} (\bibinfo {year} {1983})}\BibitemShut {NoStop}%
\bibitem [{\citenamefont {Lyth}(1997)}]{Lyth:1996im}%
  \BibitemOpen
  \bibfield  {author} {\bibinfo {author} {\bibfnamefont {D.~H.}\ \bibnamefont
  {Lyth}},\ }\Doi {10.1103/PhysRevLett.78.1861} {\bibfield  {journal} {\bibinfo
   {journal} {Phys. Rev. Lett.},\ }\textbf {\bibinfo {volume} {78}},\ \bibinfo
  {pages} {1861} (\bibinfo {year} {1997})},\ \Eprint
  {http://arxiv.org/abs/hep-ph/9606387} {arXiv:hep-ph/9606387} \BibitemShut
  {NoStop}%
\bibitem [{\citenamefont {Efstathiou}\ and\ \citenamefont
  {Mack}(2005)}]{Efstathiou:2005tq}%
  \BibitemOpen
  \bibfield  {author} {\bibinfo {author} {\bibfnamefont {G.}~\bibnamefont
  {Efstathiou}}\ and\ \bibinfo {author} {\bibfnamefont {K.~J.}\ \bibnamefont
  {Mack}},\ }\Doi {10.1088/1475-7516/2005/05/008} {\bibfield  {journal}
  {\bibinfo  {journal} {JCAP},\ }\textbf {\bibinfo {volume} {0505}},\ \bibinfo
  {pages} {008} (\bibinfo {year} {2005})},\ \Eprint
  {http://arxiv.org/abs/astro-ph/0503360} {arXiv:astro-ph/0503360} \BibitemShut
  {NoStop}%
\bibitem [{\citenamefont {Dimopoulos}\ \emph {et~al.}(1981)\citenamefont
  {Dimopoulos}, \citenamefont {Raby},\ and\ \citenamefont
  {Wilczek}}]{Dimopoulos:1981yj}%
  \BibitemOpen
  \bibfield  {author} {\bibinfo {author} {\bibfnamefont {S.}~\bibnamefont
  {Dimopoulos}}, \bibinfo {author} {\bibfnamefont {S.}~\bibnamefont {Raby}}, \
  and\ \bibinfo {author} {\bibfnamefont {F.}~\bibnamefont {Wilczek}},\ }\Doi
  {10.1103/PhysRevD.24.1681} {\bibfield  {journal} {\bibinfo  {journal} {Phys.
  Rev.},\ }\textbf {\bibinfo {volume} {D24}},\ \bibinfo {pages} {1681}
  (\bibinfo {year} {1981})}\BibitemShut {NoStop}%
\bibitem [{\citenamefont {Amaldi}\ \emph {et~al.}(1991)\citenamefont {Amaldi},
  \citenamefont {de~Boer},\ and\ \citenamefont {Furstenau}}]{Amaldi:1991cn}%
  \BibitemOpen
  \bibfield  {author} {\bibinfo {author} {\bibfnamefont {U.}~\bibnamefont
  {Amaldi}}, \bibinfo {author} {\bibfnamefont {W.}~\bibnamefont {de~Boer}}, \
  and\ \bibinfo {author} {\bibfnamefont {H.}~\bibnamefont {Furstenau}},\ }\Doi
  {10.1016/0370-2693(91)91641-8} {\bibfield  {journal} {\bibinfo  {journal}
  {Phys.Lett.},\ }\textbf {\bibinfo {volume} {B260}},\ \bibinfo {pages} {447}
  (\bibinfo {year} {1991})}\BibitemShut {NoStop}%
\bibitem [{\citenamefont {Langacker}\ and\ \citenamefont
  {Polonsky}(1993)}]{Langacker:1992rq}%
  \BibitemOpen
  \bibfield  {author} {\bibinfo {author} {\bibfnamefont {P.}~\bibnamefont
  {Langacker}}\ and\ \bibinfo {author} {\bibfnamefont {N.}~\bibnamefont
  {Polonsky}},\ }\Doi {10.1103/PhysRevD.47.4028} {\bibfield  {journal}
  {\bibinfo  {journal} {Phys.Rev.},\ }\textbf {\bibinfo {volume} {D47}},\
  \bibinfo {pages} {4028} (\bibinfo {year} {1993})},\ \Eprint
  {http://arxiv.org/abs/hep-ph/9210235} {arXiv:hep-ph/9210235 [hep-ph]}
  \BibitemShut {NoStop}%
\bibitem [{\citenamefont {Smolin}(1980)}]{Smolin:1979ca}%
  \BibitemOpen
  \bibfield  {author} {\bibinfo {author} {\bibfnamefont {L.}~\bibnamefont
  {Smolin}},\ }\Doi {10.1016/0370-2693(80)90103-3} {\bibfield  {journal}
  {\bibinfo  {journal} {Phys. Lett.},\ }\textbf {\bibinfo {volume} {B93}},\
  \bibinfo {pages} {95} (\bibinfo {year} {1980})}\BibitemShut {NoStop}%
\bibitem [{\citenamefont {Linde}(1988)}]{Linde:1987yb}%
  \BibitemOpen
  \bibfield  {author} {\bibinfo {author} {\bibfnamefont {A.~D.}\ \bibnamefont
  {Linde}},\ }\Doi {10.1016/0370-2693(88)90006-8} {\bibfield  {journal}
  {\bibinfo  {journal} {Phys. Lett.},\ }\textbf {\bibinfo {volume} {B202}},\
  \bibinfo {pages} {194} (\bibinfo {year} {1988})}\BibitemShut {NoStop}%
\bibitem [{\citenamefont {Enqvist}\ and\ \citenamefont
  {Maalampi}(1986)}]{Enqvist:1986vd}%
  \BibitemOpen
  \bibfield  {author} {\bibinfo {author} {\bibfnamefont {K.}~\bibnamefont
  {Enqvist}}\ and\ \bibinfo {author} {\bibfnamefont {J.}~\bibnamefont
  {Maalampi}},\ }\Doi {10.1016/0370-2693(86)90125-5} {\bibfield  {journal}
  {\bibinfo  {journal} {Phys. Lett.},\ }\textbf {\bibinfo {volume} {B180}},\
  \bibinfo {pages} {14} (\bibinfo {year} {1986})}\BibitemShut {NoStop}%
\bibitem [{\citenamefont {Kaloper}\ \emph {et~al.}(2011)\citenamefont
  {Kaloper}, \citenamefont {Lawrence},\ and\ \citenamefont
  {Sorbo}}]{Kaloper:2011jz}%
  \BibitemOpen
  \bibfield  {author} {\bibinfo {author} {\bibfnamefont {N.}~\bibnamefont
  {Kaloper}}, \bibinfo {author} {\bibfnamefont {A.}~\bibnamefont {Lawrence}}, \
  and\ \bibinfo {author} {\bibfnamefont {L.}~\bibnamefont {Sorbo}},\ }\Doi
  {10.1088/1475-7516/2011/03/023} {\bibfield  {journal} {\bibinfo  {journal}
  {JCAP},\ }\textbf {\bibinfo {volume} {1103}},\ \bibinfo {pages} {023}
  (\bibinfo {year} {2011})},\ \Eprint {http://arxiv.org/abs/1101.0026}
  {arXiv:1101.0026 [hep-th]} \BibitemShut {NoStop}%
\bibitem [{\citenamefont {Freese}\ \emph {et~al.}(1990)\citenamefont {Freese},
  \citenamefont {Frieman},\ and\ \citenamefont {Olinto}}]{Freese:1990rb}%
  \BibitemOpen
  \bibfield  {author} {\bibinfo {author} {\bibfnamefont {K.}~\bibnamefont
  {Freese}}, \bibinfo {author} {\bibfnamefont {J.~A.}\ \bibnamefont {Frieman}},
  \ and\ \bibinfo {author} {\bibfnamefont {A.~V.}\ \bibnamefont {Olinto}},\
  }\Doi {10.1103/PhysRevLett.65.3233} {\bibfield  {journal} {\bibinfo
  {journal} {Phys. Rev. Lett.},\ }\textbf {\bibinfo {volume} {65}},\ \bibinfo
  {pages} {3233} (\bibinfo {year} {1990})}\BibitemShut {NoStop}%
\bibitem [{\citenamefont {Arkani-Hamed}\ \emph
  {et~al.}(2003){\natexlab{a}}\citenamefont {Arkani-Hamed}, \citenamefont
  {Cheng}, \citenamefont {Creminelli},\ and\ \citenamefont
  {Randall}}]{ArkaniHamed:2003wu}%
  \BibitemOpen
  \bibfield  {author} {\bibinfo {author} {\bibfnamefont {N.}~\bibnamefont
  {Arkani-Hamed}}, \bibinfo {author} {\bibfnamefont {H.-C.}\ \bibnamefont
  {Cheng}}, \bibinfo {author} {\bibfnamefont {P.}~\bibnamefont {Creminelli}}, \
  and\ \bibinfo {author} {\bibfnamefont {L.}~\bibnamefont {Randall}},\ }\Doi
  {10.1103/PhysRevLett.90.221302} {\bibfield  {journal} {\bibinfo  {journal}
  {Phys. Rev. Lett.},\ }\textbf {\bibinfo {volume} {90}},\ \bibinfo {pages}
  {221302} (\bibinfo {year} {2003}{\natexlab{a}})},\ \Eprint
  {http://arxiv.org/abs/hep-th/0301218} {arXiv:hep-th/0301218} \BibitemShut
  {NoStop}%
\bibitem [{\citenamefont {Arkani-Hamed}\ \emph
  {et~al.}(2003){\natexlab{b}}\citenamefont {Arkani-Hamed}, \citenamefont
  {Cheng}, \citenamefont {Creminelli},\ and\ \citenamefont
  {Randall}}]{ArkaniHamed:2003mz}%
  \BibitemOpen
  \bibfield  {author} {\bibinfo {author} {\bibfnamefont {N.}~\bibnamefont
  {Arkani-Hamed}}, \bibinfo {author} {\bibfnamefont {H.-C.}\ \bibnamefont
  {Cheng}}, \bibinfo {author} {\bibfnamefont {P.}~\bibnamefont {Creminelli}}, \
  and\ \bibinfo {author} {\bibfnamefont {L.}~\bibnamefont {Randall}},\ }\Doi
  {10.1088/1475-7516/2003/07/003} {\bibfield  {journal} {\bibinfo  {journal}
  {JCAP},\ }\textbf {\bibinfo {volume} {0307}},\ \bibinfo {pages} {003}
  (\bibinfo {year} {2003}{\natexlab{b}})},\ \Eprint
  {http://arxiv.org/abs/hep-th/0302034} {arXiv:hep-th/0302034} \BibitemShut
  {NoStop}%
\bibitem [{\citenamefont {Kallosh}\ \emph {et~al.}(1995)\citenamefont
  {Kallosh}, \citenamefont {Linde}, \citenamefont {Linde},\ and\ \citenamefont
  {Susskind}}]{Kallosh:1995hi}%
  \BibitemOpen
  \bibfield  {author} {\bibinfo {author} {\bibfnamefont {R.}~\bibnamefont
  {Kallosh}}, \bibinfo {author} {\bibfnamefont {A.~D.}\ \bibnamefont {Linde}},
  \bibinfo {author} {\bibfnamefont {D.~A.}\ \bibnamefont {Linde}}, \ and\
  \bibinfo {author} {\bibfnamefont {L.}~\bibnamefont {Susskind}},\ }\Doi
  {10.1103/PhysRevD.52.912} {\bibfield  {journal} {\bibinfo  {journal} {Phys.
  Rev.},\ }\textbf {\bibinfo {volume} {D52}},\ \bibinfo {pages} {912} (\bibinfo
  {year} {1995})},\ \Eprint {http://arxiv.org/abs/hep-th/9502069}
  {arXiv:hep-th/9502069} \BibitemShut {NoStop}%
\bibitem [{\citenamefont {Banks}\ \emph {et~al.}(2003)\citenamefont {Banks},
  \citenamefont {Dine}, \citenamefont {Fox},\ and\ \citenamefont
  {Gorbatov}}]{Banks:2003sx}%
  \BibitemOpen
  \bibfield  {author} {\bibinfo {author} {\bibfnamefont {T.}~\bibnamefont
  {Banks}}, \bibinfo {author} {\bibfnamefont {M.}~\bibnamefont {Dine}},
  \bibinfo {author} {\bibfnamefont {P.~J.}\ \bibnamefont {Fox}}, \ and\
  \bibinfo {author} {\bibfnamefont {E.}~\bibnamefont {Gorbatov}},\ }\Doi
  {10.1088/1475-7516/2003/06/001} {\bibfield  {journal} {\bibinfo  {journal}
  {JCAP},\ }\textbf {\bibinfo {volume} {0306}},\ \bibinfo {pages} {001}
  (\bibinfo {year} {2003})},\ \Eprint {http://arxiv.org/abs/hep-th/0303252}
  {arXiv:hep-th/0303252} \BibitemShut {NoStop}%
\bibitem [{\citenamefont {Arkani-Hamed}\ \emph {et~al.}(2007)\citenamefont
  {Arkani-Hamed}, \citenamefont {Motl}, \citenamefont {Nicolis},\ and\
  \citenamefont {Vafa}}]{ArkaniHamed:2006dz}%
  \BibitemOpen
  \bibfield  {author} {\bibinfo {author} {\bibfnamefont {N.}~\bibnamefont
  {Arkani-Hamed}}, \bibinfo {author} {\bibfnamefont {L.}~\bibnamefont {Motl}},
  \bibinfo {author} {\bibfnamefont {A.}~\bibnamefont {Nicolis}}, \ and\
  \bibinfo {author} {\bibfnamefont {C.}~\bibnamefont {Vafa}},\ }\href@noop {}
  {\bibfield  {journal} {\bibinfo  {journal} {JHEP},\ }\textbf {\bibinfo
  {volume} {06}},\ \bibinfo {pages} {060} (\bibinfo {year} {2007})},\ \Eprint
  {http://arxiv.org/abs/hep-th/0601001} {arXiv:hep-th/0601001} \BibitemShut
  {NoStop}%
\bibitem [{\citenamefont {Silverstein}\ and\ \citenamefont
  {Westphal}(2008)}]{Silverstein:2008sg}%
  \BibitemOpen
  \bibfield  {author} {\bibinfo {author} {\bibfnamefont {E.}~\bibnamefont
  {Silverstein}}\ and\ \bibinfo {author} {\bibfnamefont {A.}~\bibnamefont
  {Westphal}},\ }\Doi {10.1103/PhysRevD.78.106003} {\bibfield  {journal}
  {\bibinfo  {journal} {Phys. Rev.},\ }\textbf {\bibinfo {volume} {D78}},\
  \bibinfo {pages} {106003} (\bibinfo {year} {2008})},\ \Eprint
  {http://arxiv.org/abs/0803.3085} {arXiv:0803.3085 [hep-th]} \BibitemShut
  {NoStop}%
\bibitem [{\citenamefont {McAllister}\ \emph {et~al.}(2010)\citenamefont
  {McAllister}, \citenamefont {Silverstein},\ and\ \citenamefont
  {Westphal}}]{McAllister:2008hb}%
  \BibitemOpen
  \bibfield  {author} {\bibinfo {author} {\bibfnamefont {L.}~\bibnamefont
  {McAllister}}, \bibinfo {author} {\bibfnamefont {E.}~\bibnamefont
  {Silverstein}}, \ and\ \bibinfo {author} {\bibfnamefont {A.}~\bibnamefont
  {Westphal}},\ }\Doi {10.1103/PhysRevD.82.046003} {\bibfield  {journal}
  {\bibinfo  {journal} {Phys.Rev.},\ }\textbf {\bibinfo {volume} {D82}},\
  \bibinfo {pages} {046003} (\bibinfo {year} {2010})},\ \Eprint
  {http://arxiv.org/abs/0808.0706} {arXiv:0808.0706 [hep-th]} \BibitemShut
  {NoStop}%
\bibitem [{\citenamefont {Kaloper}\ and\ \citenamefont
  {Sorbo}(2009){\natexlab{a}}}]{Kaloper:2008fb}%
  \BibitemOpen
  \bibfield  {author} {\bibinfo {author} {\bibfnamefont {N.}~\bibnamefont
  {Kaloper}}\ and\ \bibinfo {author} {\bibfnamefont {L.}~\bibnamefont
  {Sorbo}},\ }\Doi {10.1103/PhysRevLett.102.121301} {\bibfield  {journal}
  {\bibinfo  {journal} {Phys. Rev. Lett.},\ }\textbf {\bibinfo {volume}
  {102}},\ \bibinfo {pages} {121301} (\bibinfo {year} {2009}{\natexlab{a}})},\
  \Eprint {http://arxiv.org/abs/0811.1989} {arXiv:0811.1989 [hep-th]}
  \BibitemShut {NoStop}%
\bibitem [{\citenamefont {Dubovsky}\ \emph {et~al.}(2012)\citenamefont
  {Dubovsky}, \citenamefont {Lawrence},\ and\ \citenamefont
  {Roberts}}]{Dubovsky:2011tu}%
  \BibitemOpen
  \bibfield  {author} {\bibinfo {author} {\bibfnamefont {S.}~\bibnamefont
  {Dubovsky}}, \bibinfo {author} {\bibfnamefont {A.}~\bibnamefont {Lawrence}},
  \ and\ \bibinfo {author} {\bibfnamefont {M.~M.}\ \bibnamefont {Roberts}},\
  }\Doi {10.1007/JHEP02(2012)053} {\bibfield  {journal} {\bibinfo  {journal}
  {JHEP},\ }\textbf {\bibinfo {volume} {1202}},\ \bibinfo {pages} {053}
  (\bibinfo {year} {2012})},\ \Eprint {http://arxiv.org/abs/1105.3740}
  {arXiv:1105.3740 [hep-th]} \BibitemShut {NoStop}%
\bibitem [{\citenamefont {Dong}\ \emph {et~al.}(2011)\citenamefont {Dong},
  \citenamefont {Horn}, \citenamefont {Silverstein},\ and\ \citenamefont
  {Westphal}}]{Dong:2010in}%
  \BibitemOpen
  \bibfield  {author} {\bibinfo {author} {\bibfnamefont {X.}~\bibnamefont
  {Dong}}, \bibinfo {author} {\bibfnamefont {B.}~\bibnamefont {Horn}}, \bibinfo
  {author} {\bibfnamefont {E.}~\bibnamefont {Silverstein}}, \ and\ \bibinfo
  {author} {\bibfnamefont {A.}~\bibnamefont {Westphal}},\ }\Doi
  {10.1103/PhysRevD.84.026011} {\bibfield  {journal} {\bibinfo  {journal}
  {Phys.Rev.},\ }\textbf {\bibinfo {volume} {D84}},\ \bibinfo {pages} {026011}
  (\bibinfo {year} {2011})},\ \Eprint {http://arxiv.org/abs/1011.4521}
  {arXiv:1011.4521 [hep-th]} \BibitemShut {NoStop}%
\bibitem [{\citenamefont {McAllister}\ \emph {et~al.}()\citenamefont
  {McAllister}, \citenamefont {Senatore}, \citenamefont {Silverstein},
  \citenamefont {Westphal},\ and\ \citenamefont {Wrase}}]{monopowers}%
  \BibitemOpen
  \bibfield  {author} {\bibinfo {author} {\bibfnamefont {L.}~\bibnamefont
  {McAllister}}, \bibinfo {author} {\bibfnamefont {L.}~\bibnamefont
  {Senatore}}, \bibinfo {author} {\bibfnamefont {E.}~\bibnamefont
  {Silverstein}}, \bibinfo {author} {\bibfnamefont {A.}~\bibnamefont
  {Westphal}}, \ and\ \bibinfo {author} {\bibfnamefont {T.}~\bibnamefont
  {Wrase}},\ }\href@noop {} {\bibinfo  {journal} {in preparation}}\BibitemShut
  {NoStop}%
\bibitem [{\citenamefont {D'Amico}\ \emph
  {et~al.}(2013){\natexlab{a}}\citenamefont {D'Amico}, \citenamefont
  {Gobbetti}, \citenamefont {Schillo},\ and\ \citenamefont
  {Kleban}}]{DAmico:2012sz}%
  \BibitemOpen
\bibfield  {journal} {  }\bibfield  {author} {\bibinfo {author} {\bibfnamefont
  {G.}~\bibnamefont {D'Amico}}, \bibinfo {author} {\bibfnamefont
  {R.}~\bibnamefont {Gobbetti}}, \bibinfo {author} {\bibfnamefont
  {M.}~\bibnamefont {Schillo}}, \ and\ \bibinfo {author} {\bibfnamefont
  {M.}~\bibnamefont {Kleban}},\ }\Doi {10.1016/j.physletb.2013.07.050}
  {\bibfield  {journal} {\bibinfo  {journal} {Phys.Lett.},\ }\textbf {\bibinfo
  {volume} {B725}},\ \bibinfo {pages} {218} (\bibinfo {year}
  {2013}{\natexlab{a}})},\ \Eprint {http://arxiv.org/abs/1211.3416}
  {arXiv:1211.3416 [hep-th]} \BibitemShut {NoStop}%
\bibitem [{\citenamefont {D'Amico}\ \emph
  {et~al.}(2013){\natexlab{b}}\citenamefont {D'Amico}, \citenamefont
  {Gobbetti}, \citenamefont {Kleban},\ and\ \citenamefont
  {Schillo}}]{DAmico:2012ji}%
  \BibitemOpen
  \bibfield  {author} {\bibinfo {author} {\bibfnamefont {G.}~\bibnamefont
  {D'Amico}}, \bibinfo {author} {\bibfnamefont {R.}~\bibnamefont {Gobbetti}},
  \bibinfo {author} {\bibfnamefont {M.}~\bibnamefont {Kleban}}, \ and\ \bibinfo
  {author} {\bibfnamefont {M.}~\bibnamefont {Schillo}},\ }\Doi
  {10.1088/1475-7516/2013/03/004} {\bibfield  {journal} {\bibinfo  {journal}
  {JCAP},\ }\textbf {\bibinfo {volume} {1303}},\ \bibinfo {pages} {004}
  (\bibinfo {year} {2013}{\natexlab{b}})},\ \Eprint
  {http://arxiv.org/abs/1211.4589} {arXiv:1211.4589 [hep-th]} \BibitemShut
  {NoStop}%
\bibitem [{\citenamefont {Kobzarev}\ \emph {et~al.}(1975)\citenamefont
  {Kobzarev}, \citenamefont {Okun},\ and\ \citenamefont
  {Voloshin}}]{Kobzarev:1974cp}%
  \BibitemOpen
  \bibfield  {author} {\bibinfo {author} {\bibfnamefont {I.~Y.}\ \bibnamefont
  {Kobzarev}}, \bibinfo {author} {\bibfnamefont {L.~B.}\ \bibnamefont {Okun}},
  \ and\ \bibinfo {author} {\bibfnamefont {M.~B.}\ \bibnamefont {Voloshin}},\
  }\href@noop {} {\bibfield  {journal} {\bibinfo  {journal} {Sov. J. Nucl.
  Phys.},\ }\textbf {\bibinfo {volume} {20}},\ \bibinfo {pages} {644} (\bibinfo
  {year} {1975})}\BibitemShut {NoStop}%
\bibitem [{\citenamefont {Coleman}(1977)}]{Coleman:1977py}%
  \BibitemOpen
  \bibfield  {author} {\bibinfo {author} {\bibfnamefont {S.~R.}\ \bibnamefont
  {Coleman}},\ }\Doi {10.1103/PhysRevD.15.2929} {\bibfield  {journal} {\bibinfo
   {journal} {Phys. Rev.},\ }\textbf {\bibinfo {volume} {D15}},\ \bibinfo
  {pages} {2929} (\bibinfo {year} {1977})}\BibitemShut {NoStop}%
\bibitem [{\citenamefont {Shlaer}(2013)}]{Shlaer:2012by}%
  \BibitemOpen
  \bibfield  {author} {\bibinfo {author} {\bibfnamefont {B.}~\bibnamefont
  {Shlaer}},\ }\Doi {10.1103/PhysRevD.88.103503} {\bibfield  {journal}
  {\bibinfo  {journal} {Phys.Rev.},\ }\textbf {\bibinfo {volume} {D88}},\
  \bibinfo {pages} {103503} (\bibinfo {year} {2013})},\ \Eprint
  {http://arxiv.org/abs/1211.4024} {arXiv:1211.4024 [hep-th]} \BibitemShut
  {NoStop}%
\bibitem [{\citenamefont {Palti}\ and\ \citenamefont
  {Weigand}()}]{Palti:2014kza}%
  \BibitemOpen
  \bibfield  {author} {\bibinfo {author} {\bibfnamefont {E.}~\bibnamefont
  {Palti}}\ and\ \bibinfo {author} {\bibfnamefont {T.}~\bibnamefont
  {Weigand}},\ }\href@noop {} {}\Eprint {http://arxiv.org/abs/1403.7507}
  {arXiv:1403.7507 [hep-th]} \BibitemShut {NoStop}%
\bibitem [{\citenamefont {Kaloper}\ \emph {et~al.}(2002)\citenamefont
  {Kaloper}, \citenamefont {Kleban}, \citenamefont {Lawrence},\ and\
  \citenamefont {Shenker}}]{Kaloper:2002uj}%
  \BibitemOpen
  \bibfield  {author} {\bibinfo {author} {\bibfnamefont {N.}~\bibnamefont
  {Kaloper}}, \bibinfo {author} {\bibfnamefont {M.}~\bibnamefont {Kleban}},
  \bibinfo {author} {\bibfnamefont {A.~E.}\ \bibnamefont {Lawrence}}, \ and\
  \bibinfo {author} {\bibfnamefont {S.}~\bibnamefont {Shenker}},\ }\Doi
  {10.1103/PhysRevD.66.123510} {\bibfield  {journal} {\bibinfo  {journal}
  {Phys. Rev.},\ }\textbf {\bibinfo {volume} {D66}},\ \bibinfo {pages} {123510}
  (\bibinfo {year} {2002})},\ \Eprint {http://arxiv.org/abs/hep-th/0201158}
  {arXiv:hep-th/0201158} \BibitemShut {NoStop}%
\bibitem [{\citenamefont {Harvey}\ \emph {et~al.}(1995)\citenamefont {Harvey},
  \citenamefont {Lowe},\ and\ \citenamefont {Strominger}}]{Harvey:1995ne}%
  \BibitemOpen
  \bibfield  {author} {\bibinfo {author} {\bibfnamefont {J.~A.}\ \bibnamefont
  {Harvey}}, \bibinfo {author} {\bibfnamefont {D.~A.}\ \bibnamefont {Lowe}}, \
  and\ \bibinfo {author} {\bibfnamefont {A.}~\bibnamefont {Strominger}},\ }\Doi
  {10.1016/0370-2693(95)01144-F} {\bibfield  {journal} {\bibinfo  {journal}
  {Phys.Lett.},\ }\textbf {\bibinfo {volume} {B362}},\ \bibinfo {pages} {65}
  (\bibinfo {year} {1995})},\ \Eprint {http://arxiv.org/abs/hep-th/9507168}
  {arXiv:hep-th/9507168 [hep-th]} \BibitemShut {NoStop}%
\bibitem [{\citenamefont {Acharya}(1996)}]{Acharya:1996ci}%
  \BibitemOpen
  \bibfield  {author} {\bibinfo {author} {\bibfnamefont {B.~S.}\ \bibnamefont
  {Acharya}},\ }\Doi {10.1016/0550-3213(96)00326-4} {\bibfield  {journal}
  {\bibinfo  {journal} {Nucl.Phys.},\ }\textbf {\bibinfo {volume} {B475}},\
  \bibinfo {pages} {579} (\bibinfo {year} {1996})},\ \Eprint
  {http://arxiv.org/abs/hep-th/9603033} {arXiv:hep-th/9603033 [hep-th]}
  \BibitemShut {NoStop}%
\bibitem [{\citenamefont {Witten}(){\natexlab{a}}}]{Witten:2001uq}%
  \BibitemOpen
  \bibfield  {author} {\bibinfo {author} {\bibfnamefont {E.}~\bibnamefont
  {Witten}},\ }\href@noop {} { ({\natexlab{a}})},\ \Eprint
  {http://arxiv.org/abs/hep-th/0108165} {arXiv:hep-th/0108165 [hep-th]}
  \BibitemShut {NoStop}%
\bibitem [{\citenamefont {Acharya}\ and\ \citenamefont
  {Witten}()}]{Acharya:2001gy}%
  \BibitemOpen
  \bibfield  {author} {\bibinfo {author} {\bibfnamefont {B.~S.}\ \bibnamefont
  {Acharya}}\ and\ \bibinfo {author} {\bibfnamefont {E.}~\bibnamefont
  {Witten}},\ }\href@noop {} {}\Eprint {http://arxiv.org/abs/hep-th/0109152}
  {arXiv:hep-th/0109152 [hep-th]} \BibitemShut {NoStop}%
\bibitem [{\citenamefont {Witten}(){\natexlab{b}}}]{Witten:2001bf}%
  \BibitemOpen
  \bibfield  {author} {\bibinfo {author} {\bibfnamefont {E.}~\bibnamefont
  {Witten}},\ }\href@noop {} { ({\natexlab{b}})},\ \Eprint
  {http://arxiv.org/abs/hep-ph/0201018} {arXiv:hep-ph/0201018 [hep-ph]}
  \BibitemShut {NoStop}%
\bibitem [{\citenamefont {Beasley}\ and\ \citenamefont
  {Witten}(2002)}]{Beasley:2002db}%
  \BibitemOpen
  \bibfield  {author} {\bibinfo {author} {\bibfnamefont {C.}~\bibnamefont
  {Beasley}}\ and\ \bibinfo {author} {\bibfnamefont {E.}~\bibnamefont
  {Witten}},\ }\Doi {10.1088/1126-6708/2002/07/046} {\bibfield  {journal}
  {\bibinfo  {journal} {JHEP},\ }\textbf {\bibinfo {volume} {0207}},\ \bibinfo
  {pages} {046} (\bibinfo {year} {2002})},\ \Eprint
  {http://arxiv.org/abs/hep-th/0203061} {arXiv:hep-th/0203061 [hep-th]}
  \BibitemShut {NoStop}%
\bibitem [{\citenamefont {Kaloper}\ and\ \citenamefont
  {Sorbo}(2009){\natexlab{b}}}]{Kaloper:2008qs}%
  \BibitemOpen
  \bibfield  {author} {\bibinfo {author} {\bibfnamefont {N.}~\bibnamefont
  {Kaloper}}\ and\ \bibinfo {author} {\bibfnamefont {L.}~\bibnamefont
  {Sorbo}},\ }\Doi {10.1103/PhysRevD.79.043528} {\bibfield  {journal} {\bibinfo
   {journal} {Phys.Rev.},\ }\textbf {\bibinfo {volume} {D79}},\ \bibinfo
  {pages} {043528} (\bibinfo {year} {2009}{\natexlab{b}})},\ \Eprint
  {http://arxiv.org/abs/0810.5346} {arXiv:0810.5346 [hep-th]} \BibitemShut
  {NoStop}%
\bibitem [{\citenamefont {Kachru}\ \emph {et~al.}(2006)\citenamefont {Kachru},
  \citenamefont {McGreevy},\ and\ \citenamefont {Svrcek}}]{Kachru:2006em}%
  \BibitemOpen
  \bibfield  {author} {\bibinfo {author} {\bibfnamefont {S.}~\bibnamefont
  {Kachru}}, \bibinfo {author} {\bibfnamefont {J.}~\bibnamefont {McGreevy}}, \
  and\ \bibinfo {author} {\bibfnamefont {P.}~\bibnamefont {Svrcek}},\
  }\href@noop {} {\bibfield  {journal} {\bibinfo  {journal} {JHEP},\ }\textbf
  {\bibinfo {volume} {0604}},\ \bibinfo {pages} {023} (\bibinfo {year}
  {2006})},\ \Eprint {http://arxiv.org/abs/hep-th/0601111}
  {arXiv:hep-th/0601111 [hep-th]} \BibitemShut {NoStop}%
\bibitem [{\citenamefont {Cicoli}\ \emph {et~al.}(2012)\citenamefont {Cicoli},
  \citenamefont {Krippendorf}, \citenamefont {Mayrhofer}, \citenamefont
  {Quevedo},\ and\ \citenamefont {Valandro}}]{Cicoli:2012vw}%
  \BibitemOpen
  \bibfield  {author} {\bibinfo {author} {\bibfnamefont {M.}~\bibnamefont
  {Cicoli}}, \bibinfo {author} {\bibfnamefont {S.}~\bibnamefont {Krippendorf}},
  \bibinfo {author} {\bibfnamefont {C.}~\bibnamefont {Mayrhofer}}, \bibinfo
  {author} {\bibfnamefont {F.}~\bibnamefont {Quevedo}}, \ and\ \bibinfo
  {author} {\bibfnamefont {R.}~\bibnamefont {Valandro}},\ }\Doi
  {10.1007/JHEP09(2012)019} {\bibfield  {journal} {\bibinfo  {journal} {JHEP},\
  }\textbf {\bibinfo {volume} {1209}},\ \bibinfo {pages} {019} (\bibinfo {year}
  {2012})},\ \Eprint {http://arxiv.org/abs/1206.5237} {arXiv:1206.5237
  [hep-th]} \BibitemShut {NoStop}%
\bibitem [{\citenamefont {Cicoli}\ \emph {et~al.}(2013)\citenamefont {Cicoli},
  \citenamefont {Krippendorf}, \citenamefont {Mayrhofer}, \citenamefont
  {Quevedo},\ and\ \citenamefont {Valandro}}]{Cicoli:2013mpa}%
  \BibitemOpen
  \bibfield  {author} {\bibinfo {author} {\bibfnamefont {M.}~\bibnamefont
  {Cicoli}}, \bibinfo {author} {\bibfnamefont {S.}~\bibnamefont {Krippendorf}},
  \bibinfo {author} {\bibfnamefont {C.}~\bibnamefont {Mayrhofer}}, \bibinfo
  {author} {\bibfnamefont {F.}~\bibnamefont {Quevedo}}, \ and\ \bibinfo
  {author} {\bibfnamefont {R.}~\bibnamefont {Valandro}},\ }\Doi
  {10.1007/JHEP07(2013)150} {\bibfield  {journal} {\bibinfo  {journal} {JHEP},\
  }\textbf {\bibinfo {volume} {1307}},\ \bibinfo {pages} {150} (\bibinfo {year}
  {2013})},\ \Eprint {http://arxiv.org/abs/1304.0022} {arXiv:1304.0022
  [hep-th]} \BibitemShut {NoStop}%
\bibitem [{\citenamefont {Cicoli}\ \emph {et~al.}()\citenamefont {Cicoli},
  \citenamefont {Klevers}, \citenamefont {Krippendorf}, \citenamefont
  {Mayrhofer}, \citenamefont {Quevedo} \emph {et~al.}}]{Cicoli:2013cha}%
  \BibitemOpen
  \bibfield  {author} {\bibinfo {author} {\bibfnamefont {M.}~\bibnamefont
  {Cicoli}}, \bibinfo {author} {\bibfnamefont {D.}~\bibnamefont {Klevers}},
  \bibinfo {author} {\bibfnamefont {S.}~\bibnamefont {Krippendorf}}, \bibinfo
  {author} {\bibfnamefont {C.}~\bibnamefont {Mayrhofer}}, \bibinfo {author}
  {\bibfnamefont {F.}~\bibnamefont {Quevedo}},  \emph {et~al.},\ }\href@noop {}
  {}\Eprint {http://arxiv.org/abs/1312.0014} {arXiv:1312.0014 [hep-th]}
  \BibitemShut {NoStop}%
\bibitem [{\citenamefont {Hall}\ \emph {et~al.}()\citenamefont {Hall},
  \citenamefont {Nomura},\ and\ \citenamefont {Shirai}}]{Hall:2014vga}%
  \BibitemOpen
  \bibfield  {author} {\bibinfo {author} {\bibfnamefont {L.~J.}\ \bibnamefont
  {Hall}}, \bibinfo {author} {\bibfnamefont {Y.}~\bibnamefont {Nomura}}, \ and\
  \bibinfo {author} {\bibfnamefont {S.}~\bibnamefont {Shirai}},\ }\href@noop {}
  {}\Eprint {http://arxiv.org/abs/1403.8138} {arXiv:1403.8138 [hep-ph]}
  \BibitemShut {NoStop}%
\bibitem [{\citenamefont {Baumann}\ \emph {et~al.}(2009)\citenamefont {Baumann}
  \emph {et~al.}}]{Baumann:2008aq}%
  \BibitemOpen
  \bibfield  {author} {\bibinfo {author} {\bibfnamefont {D.}~\bibnamefont
  {Baumann}} \emph {et~al.} (\bibinfo {collaboration} {CMBPol Study Team}),\
  }\Doi {10.1063/1.3160885} {\bibfield  {journal} {\bibinfo  {journal} {AIP
  Conf.Proc.},\ }\textbf {\bibinfo {volume} {1141}},\ \bibinfo {pages} {10}
  (\bibinfo {year} {2009})},\ \Eprint {http://arxiv.org/abs/0811.3919}
  {arXiv:0811.3919 [astro-ph]} \BibitemShut {NoStop}%
\bibitem [{\citenamefont {Creminelli}\ \emph {et~al.}()\citenamefont
  {Creminelli}, \citenamefont {Nacir}, \citenamefont {Simonovi{\'c}},
  \citenamefont {Trevisan},\ and\ \citenamefont
  {Zaldarriaga}}]{Creminelli:2014oaa}%
  \BibitemOpen
  \bibfield  {author} {\bibinfo {author} {\bibfnamefont {P.}~\bibnamefont
  {Creminelli}}, \bibinfo {author} {\bibfnamefont {D.~L.}\ \bibnamefont
  {Nacir}}, \bibinfo {author} {\bibfnamefont {M.}~\bibnamefont
  {Simonovi{\'c}}}, \bibinfo {author} {\bibfnamefont {G.}~\bibnamefont
  {Trevisan}}, \ and\ \bibinfo {author} {\bibfnamefont {M.}~\bibnamefont
  {Zaldarriaga}},\ }\href@noop {} {}\Eprint {http://arxiv.org/abs/1404.1065}
  {arXiv:1404.1065 [astro-ph.CO]} \BibitemShut {NoStop}%
\bibitem [{\citenamefont {Freivogel}\ \emph {et~al.}(2006)\citenamefont
  {Freivogel}, \citenamefont {Kleban}, \citenamefont {Rodriguez~Martinez},\
  and\ \citenamefont {Susskind}}]{Freivogel:2005vv}%
  \BibitemOpen
  \bibfield  {author} {\bibinfo {author} {\bibfnamefont {B.}~\bibnamefont
  {Freivogel}}, \bibinfo {author} {\bibfnamefont {M.}~\bibnamefont {Kleban}},
  \bibinfo {author} {\bibfnamefont {M.}~\bibnamefont {Rodriguez~Martinez}}, \
  and\ \bibinfo {author} {\bibfnamefont {L.}~\bibnamefont {Susskind}},\ }\Doi
  {10.1088/1126-6708/2006/03/039} {\bibfield  {journal} {\bibinfo  {journal}
  {JHEP},\ }\textbf {\bibinfo {volume} {03}},\ \bibinfo {pages} {039} (\bibinfo
  {year} {2006})},\ \Eprint {http://arxiv.org/abs/hep-th/0505232}
  {arXiv:hep-th/0505232} \BibitemShut {NoStop}%
\bibitem [{\citenamefont {Freivogel}\ \emph {et~al.}()\citenamefont
  {Freivogel}, \citenamefont {Kleban}, \citenamefont {Martinez},\ and\
  \citenamefont {Susskind}}]{mattlenny}%
  \BibitemOpen
  \bibfield  {author} {\bibinfo {author} {\bibfnamefont {B.}~\bibnamefont
  {Freivogel}}, \bibinfo {author} {\bibfnamefont {M.}~\bibnamefont {Kleban}},
  \bibinfo {author} {\bibfnamefont {M.~R.}\ \bibnamefont {Martinez}}, \ and\
  \bibinfo {author} {\bibfnamefont {L.}~\bibnamefont {Susskind}},\ }\href@noop
  {} {}\Eprint {http://arxiv.org/abs/1404.2274} {arXiv:1404.2274 [astro-ph.CO]}
  \BibitemShut {NoStop}%
\bibitem [{\citenamefont {D'Amico}\ \emph
  {et~al.}(2013){\natexlab{c}}\citenamefont {D'Amico}, \citenamefont
  {Gobbetti}, \citenamefont {Kleban},\ and\ \citenamefont
  {Schillo}}]{DAmico:2013iaa}%
  \BibitemOpen
  \bibfield  {author} {\bibinfo {author} {\bibfnamefont {G.}~\bibnamefont
  {D'Amico}}, \bibinfo {author} {\bibfnamefont {R.}~\bibnamefont {Gobbetti}},
  \bibinfo {author} {\bibfnamefont {M.}~\bibnamefont {Kleban}}, \ and\ \bibinfo
  {author} {\bibfnamefont {M.}~\bibnamefont {Schillo}},\ }\Doi
  {10.1088/1475-7516/2013/11/013} {\bibfield  {journal} {\bibinfo  {journal}
  {JCAP},\ }\textbf {\bibinfo {volume} {1311}},\ \bibinfo {pages} {013}
  (\bibinfo {year} {2013}{\natexlab{c}})},\ \Eprint
  {http://arxiv.org/abs/1306.6872} {arXiv:1306.6872 [astro-ph.CO]} \BibitemShut
  {NoStop}%
\bibitem [{\citenamefont {Eriksen}\ \emph {et~al.}(2004)\citenamefont
  {Eriksen}, \citenamefont {Hansen}, \citenamefont {Banday}, \citenamefont
  {Gorski},\ and\ \citenamefont {Lilje}}]{Eriksen:2003db}%
  \BibitemOpen
  \bibfield  {author} {\bibinfo {author} {\bibfnamefont {H.}~\bibnamefont
  {Eriksen}}, \bibinfo {author} {\bibfnamefont {F.}~\bibnamefont {Hansen}},
  \bibinfo {author} {\bibfnamefont {A.}~\bibnamefont {Banday}}, \bibinfo
  {author} {\bibfnamefont {K.}~\bibnamefont {Gorski}}, \ and\ \bibinfo {author}
  {\bibfnamefont {P.}~\bibnamefont {Lilje}},\ }\Doi {10.1086/382267} {\bibfield
   {journal} {\bibinfo  {journal} {Astrophys.J.},\ }\textbf {\bibinfo {volume}
  {605}},\ \bibinfo {pages} {14} (\bibinfo {year} {2004})},\ \Eprint
  {http://arxiv.org/abs/astro-ph/0307507} {arXiv:astro-ph/0307507 [astro-ph]}
  \BibitemShut {NoStop}%
\bibitem [{\citenamefont {Sloth}()}]{sloth}%
  \BibitemOpen
  \bibfield  {author} {\bibinfo {author} {\bibfnamefont {M.}~\bibnamefont
  {Sloth}},\ }\href@noop {} {}\Eprint {http://arxiv.org/abs/1403.8051}
  {arXiv:1403.8051 [hep-th]} \BibitemShut {NoStop}%
\bibitem [{\citenamefont {Liddle}\ and\ \citenamefont
  {Lyth}(1992)}]{Liddle:1992wi}%
  \BibitemOpen
  \bibfield  {author} {\bibinfo {author} {\bibfnamefont {A.~R.}\ \bibnamefont
  {Liddle}}\ and\ \bibinfo {author} {\bibfnamefont {D.~H.}\ \bibnamefont
  {Lyth}},\ }\Doi {10.1016/0370-2693(92)91393-N} {\bibfield  {journal}
  {\bibinfo  {journal} {Phys.Lett.},\ }\textbf {\bibinfo {volume} {B291}},\
  \bibinfo {pages} {391} (\bibinfo {year} {1992})},\ \Eprint
  {http://arxiv.org/abs/astro-ph/9208007} {arXiv:astro-ph/9208007 [astro-ph]}
  \BibitemShut {NoStop}%
\bibitem [{\citenamefont {Dodelson}()}]{Dodelson:2014exa}%
  \BibitemOpen
  \bibfield  {author} {\bibinfo {author} {\bibfnamefont {S.}~\bibnamefont
  {Dodelson}},\ }\href@noop {} {}\Eprint {http://arxiv.org/abs/1403.6310}
  {arXiv:1403.6310 [astro-ph.CO]} \BibitemShut {NoStop}%
\bibitem [{\citenamefont {Kobayashi}\ and\ \citenamefont
  {Seto}()}]{Kobayashi:2014jga}%
  \BibitemOpen
  \bibfield  {author} {\bibinfo {author} {\bibfnamefont {T.}~\bibnamefont
  {Kobayashi}}\ and\ \bibinfo {author} {\bibfnamefont {O.}~\bibnamefont
  {Seto}},\ }\href@noop {} {}\Eprint {http://arxiv.org/abs/1403.5055}
  {arXiv:1403.5055 [astro-ph.CO]} \BibitemShut {NoStop}%
\bibitem [{\citenamefont {Collins}\ \emph {et~al.}()\citenamefont {Collins},
  \citenamefont {Holman},\ and\ \citenamefont {Vardanyan}}]{Collins:2014yua}%
  \BibitemOpen
  \bibfield  {author} {\bibinfo {author} {\bibfnamefont {H.}~\bibnamefont
  {Collins}}, \bibinfo {author} {\bibfnamefont {R.}~\bibnamefont {Holman}}, \
  and\ \bibinfo {author} {\bibfnamefont {T.}~\bibnamefont {Vardanyan}},\
  }\href@noop {} {}\Eprint {http://arxiv.org/abs/1403.4592} {arXiv:1403.4592
  [hep-th]} \BibitemShut {NoStop}%
\bibitem [{\citenamefont {Contaldi}\ \emph {et~al.}()\citenamefont {Contaldi},
  \citenamefont {Peloso},\ and\ \citenamefont {Sorbo}}]{Contaldi:2014zua}%
  \BibitemOpen
  \bibfield  {author} {\bibinfo {author} {\bibfnamefont {C.~R.}\ \bibnamefont
  {Contaldi}}, \bibinfo {author} {\bibfnamefont {M.}~\bibnamefont {Peloso}}, \
  and\ \bibinfo {author} {\bibfnamefont {L.}~\bibnamefont {Sorbo}},\
  }\href@noop {} {}\Eprint {http://arxiv.org/abs/1403.4596} {arXiv:1403.4596
  [astro-ph.CO]} \BibitemShut {NoStop}%
\bibitem [{\citenamefont {Kehagias}\ and\ \citenamefont
  {Riotto}()}]{Kehagias:2014wza}%
  \BibitemOpen
  \bibfield  {author} {\bibinfo {author} {\bibfnamefont {A.}~\bibnamefont
  {Kehagias}}\ and\ \bibinfo {author} {\bibfnamefont {A.}~\bibnamefont
  {Riotto}},\ }\href@noop {} {}\Eprint {http://arxiv.org/abs/1403.4811}
  {arXiv:1403.4811 [astro-ph.CO]} \BibitemShut {NoStop}%
\bibitem [{\citenamefont {Dent}\ \emph {et~al.}()\citenamefont {Dent},
  \citenamefont {Krauss},\ and\ \citenamefont {Mathur}}]{Dent:2014rga}%
  \BibitemOpen
  \bibfield  {author} {\bibinfo {author} {\bibfnamefont {J.~B.}\ \bibnamefont
  {Dent}}, \bibinfo {author} {\bibfnamefont {L.~M.}\ \bibnamefont {Krauss}}, \
  and\ \bibinfo {author} {\bibfnamefont {H.}~\bibnamefont {Mathur}},\
  }\href@noop {} {}\Eprint {http://arxiv.org/abs/1403.5166} {arXiv:1403.5166
  [astro-ph.CO]} \BibitemShut {NoStop}%
\bibitem [{\citenamefont {Freese}\ and\ \citenamefont
  {Kinney}()}]{Freese:2014nla}%
  \BibitemOpen
  \bibfield  {author} {\bibinfo {author} {\bibfnamefont {K.}~\bibnamefont
  {Freese}}\ and\ \bibinfo {author} {\bibfnamefont {W.~H.}\ \bibnamefont
  {Kinney}},\ }\href@noop {} {}\Eprint {http://arxiv.org/abs/1403.5277}
  {arXiv:1403.5277 [astro-ph.CO]} \BibitemShut {NoStop}%
\bibitem [{\citenamefont {Czerny}\ \emph {et~al.}()\citenamefont {Czerny},
  \citenamefont {Higaki},\ and\ \citenamefont {Takahashi}}]{Czerny:2014qqa}%
  \BibitemOpen
  \bibfield  {author} {\bibinfo {author} {\bibfnamefont {M.}~\bibnamefont
  {Czerny}}, \bibinfo {author} {\bibfnamefont {T.}~\bibnamefont {Higaki}}, \
  and\ \bibinfo {author} {\bibfnamefont {F.}~\bibnamefont {Takahashi}},\
  }\href@noop {} {}\Eprint {http://arxiv.org/abs/1403.5883} {arXiv:1403.5883
  [hep-ph]} \BibitemShut {NoStop}%
\bibitem [{\citenamefont {Lyth}()}]{Lyth:2014yya}%
  \BibitemOpen
  \bibfield  {author} {\bibinfo {author} {\bibfnamefont {D.~H.}\ \bibnamefont
  {Lyth}},\ }\href@noop {} {}\Eprint {http://arxiv.org/abs/1403.7323}
  {arXiv:1403.7323 [hep-ph]} \BibitemShut {NoStop}%
\bibitem [{\citenamefont {Di~Bari}\ \emph {et~al.}()\citenamefont {Di~Bari},
  \citenamefont {King}, \citenamefont {Luhn}, \citenamefont {Merle},\ and\
  \citenamefont {Schmidt-May}}]{DiBari:2014oja}%
  \BibitemOpen
  \bibfield  {author} {\bibinfo {author} {\bibfnamefont {P.}~\bibnamefont
  {Di~Bari}}, \bibinfo {author} {\bibfnamefont {S.~F.}\ \bibnamefont {King}},
  \bibinfo {author} {\bibfnamefont {C.}~\bibnamefont {Luhn}}, \bibinfo {author}
  {\bibfnamefont {A.}~\bibnamefont {Merle}}, \ and\ \bibinfo {author}
  {\bibfnamefont {A.}~\bibnamefont {Schmidt-May}},\ }\href@noop {} {}\Eprint
  {http://arxiv.org/abs/1404.0009} {arXiv:1404.0009 [hep-ph]} \BibitemShut
  {NoStop}%
\bibitem [{\citenamefont {Ashoorioon}\ \emph {et~al.}()\citenamefont
  {Ashoorioon}, \citenamefont {Dimopoulos}, \citenamefont {Sheikh-Jabbari},\
  and\ \citenamefont {Shiu}}]{Ashoorioon:2014nta}%
  \BibitemOpen
  \bibfield  {author} {\bibinfo {author} {\bibfnamefont {A.}~\bibnamefont
  {Ashoorioon}}, \bibinfo {author} {\bibfnamefont {K.}~\bibnamefont
  {Dimopoulos}}, \bibinfo {author} {\bibfnamefont {M.}~\bibnamefont
  {Sheikh-Jabbari}}, \ and\ \bibinfo {author} {\bibfnamefont {G.}~\bibnamefont
  {Shiu}},\ }\href@noop {} {}\Eprint {http://arxiv.org/abs/1403.6099}
  {arXiv:1403.6099 [hep-th]} \BibitemShut {NoStop}%
\bibitem [{\citenamefont {Ho}\ and\ \citenamefont {Hsu}()}]{Ho:2014xza}%
  \BibitemOpen
  \bibfield  {author} {\bibinfo {author} {\bibfnamefont {C.~M.}\ \bibnamefont
  {Ho}}\ and\ \bibinfo {author} {\bibfnamefont {S.~D.~H.}\ \bibnamefont
  {Hsu}},\ }\href@noop {} {}\Eprint {http://arxiv.org/abs/1404.0745}
  {arXiv:1404.0745 [hep-ph]} \BibitemShut {NoStop}%
\bibitem [{\citenamefont {Marchesano}\ \emph {et~al.}()\citenamefont
  {Marchesano}, \citenamefont {Shiu},\ and\ \citenamefont
  {Uranga}}]{Marchesano:2014mla}%
  \BibitemOpen
  \bibfield  {author} {\bibinfo {author} {\bibfnamefont {F.}~\bibnamefont
  {Marchesano}}, \bibinfo {author} {\bibfnamefont {G.}~\bibnamefont {Shiu}}, \
  and\ \bibinfo {author} {\bibfnamefont {A.~M.}\ \bibnamefont {Uranga}},\
  }\href@noop {} {}\Eprint {http://arxiv.org/abs/1404.3040} {arXiv:1404.3040
  [hep-th]} \BibitemShut {NoStop}%
\bibitem [{\citenamefont {Hebecker}\ \emph {et~al.}()\citenamefont {Hebecker},
  \citenamefont {Kraus},\ and\ \citenamefont {Witkowski}}]{Hebecker:2014eua}%
  \BibitemOpen
  \bibfield  {author} {\bibinfo {author} {\bibfnamefont {A.}~\bibnamefont
  {Hebecker}}, \bibinfo {author} {\bibfnamefont {S.~C.}\ \bibnamefont {Kraus}},
  \ and\ \bibinfo {author} {\bibfnamefont {L.~T.}\ \bibnamefont {Witkowski}},\
  }\href@noop {} {}\Eprint {http://arxiv.org/abs/1404.3711} {arXiv:1404.3711
  [hep-th]} \BibitemShut {NoStop}%
\bibitem [{\citenamefont {Blumenhagen}\ and\ \citenamefont
  {Plauschinn}()}]{Blumenhagen:2014gta}%
  \BibitemOpen
  \bibfield  {author} {\bibinfo {author} {\bibfnamefont {R.}~\bibnamefont
  {Blumenhagen}}\ and\ \bibinfo {author} {\bibfnamefont {E.}~\bibnamefont
  {Plauschinn}},\ }\href@noop {} {}\Eprint {http://arxiv.org/abs/1404.3542}
  {arXiv:1404.3542 [hep-th]} \BibitemShut {NoStop}%
\bibitem [{\citenamefont {Harigaya}\ and\ \citenamefont
  {Ibe}()}]{Harigaya:2014eta}%
  \BibitemOpen
  \bibfield  {author} {\bibinfo {author} {\bibfnamefont {K.}~\bibnamefont
  {Harigaya}}\ and\ \bibinfo {author} {\bibfnamefont {M.}~\bibnamefont {Ibe}},\
  }\href@noop {} {}\Eprint {http://arxiv.org/abs/1404.3511} {arXiv:1404.3511
  [hep-ph]} \BibitemShut {NoStop}%
\bibitem [{\citenamefont {Grimm}()}]{Grimm:2014vva}%
  \BibitemOpen
  \bibfield  {author} {\bibinfo {author} {\bibfnamefont {T.~W.}\ \bibnamefont
  {Grimm}},\ }\href@noop {} {}\Eprint {http://arxiv.org/abs/1404.4268}
  {arXiv:1404.4268 [hep-th]} \BibitemShut {NoStop}%
\bibitem [{\citenamefont {Ibanez}\ and\ \citenamefont
  {Valenzuela}()}]{Ibanez:2014kia}%
  \BibitemOpen
  \bibfield  {author} {\bibinfo {author} {\bibfnamefont {L.~E.}\ \bibnamefont
  {Ibanez}}\ and\ \bibinfo {author} {\bibfnamefont {I.}~\bibnamefont
  {Valenzuela}},\ }\href@noop {} {}\Eprint {http://arxiv.org/abs/1404.5235}
  {arXiv:1404.5235 [hep-th]} \BibitemShut {NoStop}%
\end{thebibliography}%

\end{document}